\documentclass[a4paper,11pt]{article}
\pdfoutput=1 % if your are submitting a pdflatex (i.e. if you have
\usepackage{jheppub} 

\usepackage{xcolor}
\usepackage{graphicx}
\usepackage{subfig}
\usepackage{multirow}
\usepackage{amsmath}
\usepackage{textcomp}
\usepackage{appendix}
\usepackage{xcolor}
\usepackage{wasysym}
\usepackage{textcomp}
\usepackage{amssymb}

\newlength{\figwidth}
\title{Optimised sensitivity to leptonic CP violation from spectral information:
the LBNO case at 2300~km baseline}

\author[o]{S.K.\,Agarwalla,}
\author[a]{L.\,Agostino,}
\author[u]{M.\,Aittola,}
\author[b]{A.\,Alekou,}
\author[x]{B.\,Andrieu,}
\author[b]{F.\,Antoniou,}
\author[aa]{R.\,Asfandiyarov,}
\author[y]{D.\,Autiero,}
\author[k]{O.\,B\'esida,}
\author[r]{A.\,Balik,}
% ...

\affiliation[a]{APC, AstroParticule et Cosmologie, Universit\'e Paris Diderot, CNRS/IN2P3, CEA/Irfu, Observatoire de Paris, Sorbonne Paris Cit\'e, 10, rue Alice Domon et L\'eonie Duquet, 75205 Paris Cedex 13, France}
\affiliation[b]{CERN, Geneva, Switzerland}
% ...
\affiliation[k]{IRFU, CEA Saclay, Gif-sur-Yvette, France}
% ...
\affiliation[o]{Institute of Physics, Sachivalaya Marg, Sainik School Post, Bhubaneswar 751005, India}
% ...
\affiliation[r]{LAPP, Universit\'e de Savoie, CNRS/IN2P3, F-74941 Annecy-le-Vieux, France}
% ...
\affiliation[u]{Oulu Southern Institute and Department of Physics, University of Oulu, Finland}
% ...
\affiliation[x]{UPMC, Universit\'e Paris Diderot, CNRS/IN2P3, Laboratoire de Physique Nucl\'eaire et de Hautes Energies (LPNHE), Paris, France}
\affiliation[y]{Universit\'e de Lyon, Universit\'e Claude Bernard Lyon 1, IPN Lyon (IN2P3), Villeurbanne, France}
% ...
\affiliation[aa]{University of Geneva, Section de Physique, DPNC, Geneva, Switzerland}
% ...

\abstract{One of the main goals of the Long Baseline Neutrino Observatory (LBNO) is to study the $L/E$ behaviour (spectral information) 
of the electron neutrino and antineutrino appearance probabilities, in order to determine the unknown CP-violation phase $\delta_{CP}$ and discover CP-violation in the leptonic sector. The result is based on the measurement of the appearance probabilities in a broad range of energies, covering the 1st and 2nd oscillation maxima, at a very long baseline of 2300~km. 
The sensitivity of the experiment can be maximised by optimising the energy spectra of the neutrino and anti-neutrino fluxes.
Such an optimisation requires  exploring an extended range of parameters describing in details the geometries and properties of the primary protons, hadron target and focusing elements in the neutrino beam line. 
In this paper we present a numerical solution that leads to an optimised energy spectra and study its impact on the sensitivity of LBNO to discover leptonic CP violation.
In the optimised flux both 1st and 2nd oscillation maxima play an important role in the CP sensitivity. The studies also show that this configuration is less sensitive to systematic errors (e.g. on the total event rates) than an experiment which mainly relies on 
the neutrino-antineutrino asymmetry at the 1st maximum to determine
the existence of CP-violation. }

\begin{document}
\maketitle

%%%%%%%%%%%%%%%%%%%%%%%%%%%%%%%%%%%%%%%%%%%%%%%%%%%%%%%%%%%%%%%%%%%%%
%             Introduction
%%%%%%%%%%%%%%%%%%%%%%%%%%%%%%%%%%%%%%%%%%%%%%%%%%%%%%%%%%%%%%%%%%%%%
\section{Introduction}
\label{sec:intro}

The proposed Long Baseline Neutrino Observatory (LBNO) \cite{Stahl:2012exa,Agarwalla::2013kaa} will seek to address fundamental questions in particle and astroparticle physics. Following the development of a conceptual report as part of LAGUNA and LAGUNA-LBNO studies~\cite{Patzak:2012rz,Rubbia:2013zqa}, LBNO consists in deploying a large double-phase liquid argon TPC~\cite{Rubbia:2009md} deep underground in the Pyh\"{a}salmi mine, Finland, profiting from the excellent  infrastructure existing there. In the first phase, a detector with a $\sim 24$ kton mass is foreseen. The total fiducial mass is then augmented to 70 kton with an addition of a second $\sim 50$ kton detector. 

The major goals of future long baseline experiments such as the proposed LBNO, LBNE~\cite{Adams:2013qkq,Bass:2013vcg} and HyperKamiokande~\cite{Kearns:2013lea} are to precisely study the neutrino flavour oscillations occurring on the path
of neutrinos through the Earth, to obtain the conclusive determination of the mass hierarchy (MH) and the search for leptonic CP-violation (CPV). 

\begin{figure}[t]
\center
\includegraphics[width=0.975\textwidth]{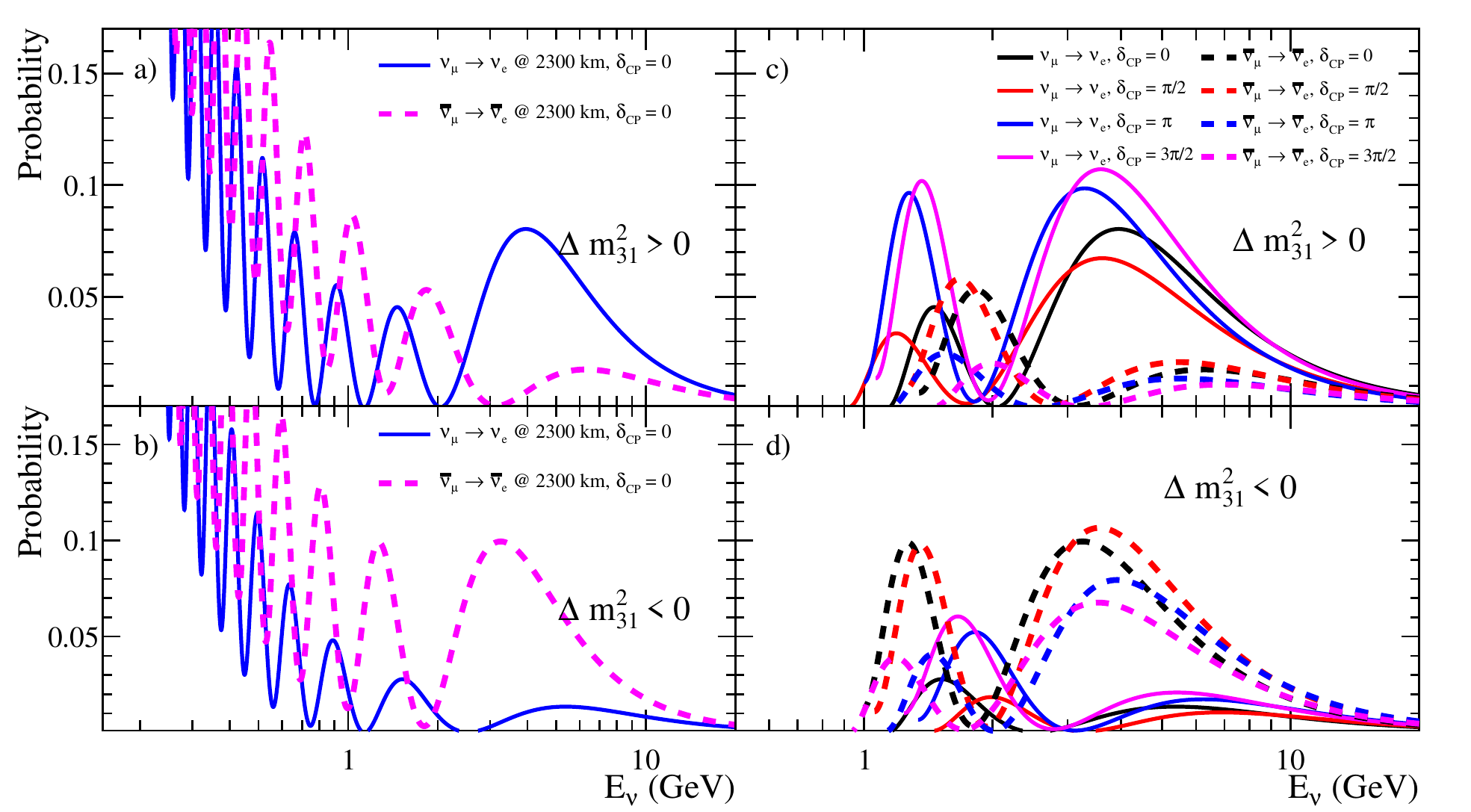}
\caption{Neutrino flavour oscillation probability at a baseline of 2300~km for 
$\nu_\mu\rightarrow \nu_e$ (resp. $\bar\nu_\mu\rightarrow \bar\nu_e$) 
as a function of the neutrino energy $E_\nu$:
a) $\Delta m^2_{31}>0$ and $\delta_{CP}=0$;
b) $\Delta m^2_{31}<0$ and $\delta_{CP}=0$;
c) $\Delta m^2_{31}>0$ and $\delta_{CP}=0, \pi/2, \pi,3\pi/2$;
d) $\Delta m^2_{31}<0$ and $\delta_{CP}=0, \pi/2, \pi,3\pi/2$
(see Table~\ref{tab:oscparam_may2014} for the other oscillation parameters assumed.)
}
\label{fig:oscprob}
\end{figure}

The neutrino flavour oscillation probabilities of 
$\nu_\mu\rightarrow \nu_e$ (resp. $\bar\nu_\mu\rightarrow \bar\nu_e$) 
as a function of the neutrino energy $E_\nu$ are illustrated in Figure~\ref{fig:oscprob} for 
a) $\Delta m^2_{31}>0$ and $\delta_{CP}=0$,
b) $\Delta m^2_{31}<0$ and $\delta_{CP}=0$,
c) $\Delta m^2_{31}>0$ and $\delta_{CP}=0, \pi/2, \pi,3\pi/2$, and
d) $\Delta m^2_{31}<0$ and $\delta_{CP}=0, \pi/2, \pi,3\pi/2$.
(see Table~\ref{tab:oscparam_may2014} for the other oscillation parameters assumed.)

For MH, a long neutrino path through Earth ($>1300$~km) induces a large asymmetry between $\nu_\mu \rightarrow \nu_e$ and $\bar{\nu}_\mu\rightarrow \bar{\nu}_e$ oscillation probabilities which depends on the sign of $\Delta m^2_{31}$ (if MH is -1(+1) (anti)neutrino oscillation are suppressed), thus allowing in principle for a fast determination of MH. This feature is important as
the knowledge of MH is a prerequisite to a full determination of CPV which can otherwise be affected 
by degeneracies that could confuse the determination of $\delta_{CP}$~\cite{Barger:2001yr}. 
As a concrete example, the studies presented in Ref.\cite{Stahl:2012exa,Agarwalla::2013kaa} considering an initial 20~kton detector
and 700~kW beam power, show that a baseline
of at least 1700~km is needed to guarantee a $5\sigma$~C.L. determination of MH over the full phase space
within 10 years of running. A baseline of $>2000$~km allows to reach this level in 2-4 years.

For CPV, a very long baseline gives access to both 1st and 2nd flavour oscillation maxima of $\nu_e$ and $\bar\nu_e$ appearance probabilities, making it possible to measure $\delta_{CP}$ from the $L/E$ behaviour of the oscillation probability over a wide energy range. 
Since the dependence of the oscillation probability on $\delta_{CP}$ is  stronger at the 2nd maximum than
at the 1st maximum, the sensitivity to CPV is greater in this $L/E$ region (see e.g.~\cite{Coloma:2011pg}).  
Therefore, one is less sensitive to the effects 
of systematic uncertainties compared to measurements focused on the 1st maximum only. Accordingly LBNO assumes
a 3\%  uncertainty on the signal normalisation and 10\% for the backgrounds. In comparison, LBNE at a baseline
of 1300~km mostly focusing on the 1st maximum assumes 1\%  uncertainty in the signal normalisation 
and 5\% for the backgrounds~\cite{Adams:2013qkq}. As pointed out in Ref.~\cite{Bass:2013vcg}, 
achieving this latter level of precisions for signal and backgrounds will 
require a well-designed near detector
and careful analysis of detector efficiencies and other systematic errors. It is fair to say
that the reality of achieving such precisions still needs to be proven.

Access to the 2nd maximum is also more easily achieved at longer baselines.
It is well known that conventional neutrino beams (with currently understood technological limits on magnetic field strengths)
are not efficient at focusing hadrons with energies below $\sim 1$~GeV~\cite{Bass:2013vcg}. 
Taking into account vanishing neutrino cross-sections at low energies (in particular for antineutrinos),
the measurement of the 2nd oscillation maximum requires in practice a baseline greater than 
$1500$~km~\cite{Agarwalla::2013kaa}. The $\nu_\mu\rightarrow \nu_\tau$~CC events where $\tau\rightarrow e\nu\nu$, 
which become more important at longer baselines due to the higher energy tails of the neutrino flux, 
act a priori as a background to electron appearance. But, as shown in Refs.~\cite{Stahl:2012exa,Agarwalla::2013kaa},
they can be kinematically separated exploiting the
excellent kinematic reconstruction of liquid argon detectors. 

In conclusion, the CERN-Pyh\"{a}salmi baseline of 2300~km studied for LBNO
offers unique ways to address the questions of MH and CPV and opens the possibility 
to tune the beam to fully exploit the spectral information over
a broad energy spectrum. 
In order to implement this strategy, the neutrino beam profile has to be optimised, so that the energy spectrum of the resultant neutrino flux is tuned to give the best experimental sensitivity to CPV for the given baseline.
This in general is a complex task that requires exploring a large phase-space of parameters,
such as for instance
target material and geometry, horn focusing elements, and decay volume dimensions. Moreover the optimisation method 
is also not unique. For example, one could attempt to maximise the neutrino yield in the energy window around the 2nd oscillation maximum. But a more balanced distribution of the neutrinos between 1st and 2nd oscillation maxima could  give a better outcome. We use a sophisticated genetic algorithm that mimics the process of natural
selection in order to find solutions that maximise the chosen figure-of-merit. Three different ``fitness criteria'' (see text)
are implemented and compared in our study.

The aim of this paper is to present the improved sensitivity to
leptonic CPV after optimisation of the target-horn focusing
system in the case of LBNO with the baseline of 2300~km. 
The numerical procedure employed in finding the optimal energy spectra of the CN2PY beamline will also be discussed.
The article is organised as follows. In Section~\ref{sec:nuflux} we introduce the options for the CERN neutrino beam and describe the procedure adopted for the optimisation of the neutrino flux. Then, in Section~\ref{sec:analysis}, we review the analysis method to calculate the experimental sensitivity to CPV. Section~\ref{sec:results} contains the results obtained with the optimised neutrino fluxes and, finally, some concluding remarks are offered in Section~\ref{sec:concl}.

%%%%%%%%%%%%%%%%%%%%%%%%%%%%%%%%%%%%%%%%%%%%%%%%%%%%%%%%%%%%%%%%%%%%%
%             Neutrino flux optimisation
%%%%%%%%%%%%%%%%%%%%%%%%%%%%%%%%%%%%%%%%%%%%%%%%%%%%%%%%%%%%%%%%%%%%%
\section{The CERN neutrino beam}
\label{sec:nuflux}

\subsection{Primary proton beam}
\label{ssec:nuflux_beam}
LBNO consists of a conventional neutrino beam produced by a high-intensity proton beam impinging on a target.
Two options for the primary proton beam are foreseen for the two successive phases of the experiment. In the first phase, an upgraded SPS will deliver a 400 GeV beam at about 700 kW beam power to the target, with an expected integrated yearly exposure of about $1.0\times 10^{20}$ protons on target (POT). In the second phase, a primary beam is foreseen to be provided by a high power PS (HPPS) facility\cite{ipac13_hpps} which will deliver a 50 GeV 2 MW proton beam and integrated yearly exposure of about $3.5\times 10^{21}$ POT. The main parameters of both beams are listed in Table~\ref{tab:myfifthtable}. 

\begin{table}[ht]
\centering
\begin{tabular}{l|c|c}
\hline
\hline
\multirow{2}{*}{Parameter}  & LBNO Phase I  & LBNO Phase II \\
                            & (20kton DLAr) & (70 kton DLAr) \\
\hline
Proton injector    & SPS & HPPS \\
Proton beam energy & 400 GeV & 50 GeV \\
Nominal beam power & 0.75 MW & 2.0 MW \\
Integrated yearly exposure & $1.0\times 10^{20}$ POT & $3.5 \times 10^{21}$ POT\\
\hline 
\hline
\end{tabular}
\caption{Primary beam parameters envisioned for the 1st and 2nd phase of LBNO.}
\label{tab:myfifthtable}
\end{table}

\subsection{Monte Carlo simulation of neutrino beam}
\label{ssec:nuflux_beammc}

\begin{figure}[t]
\center
\includegraphics[width=\the\figwidth]{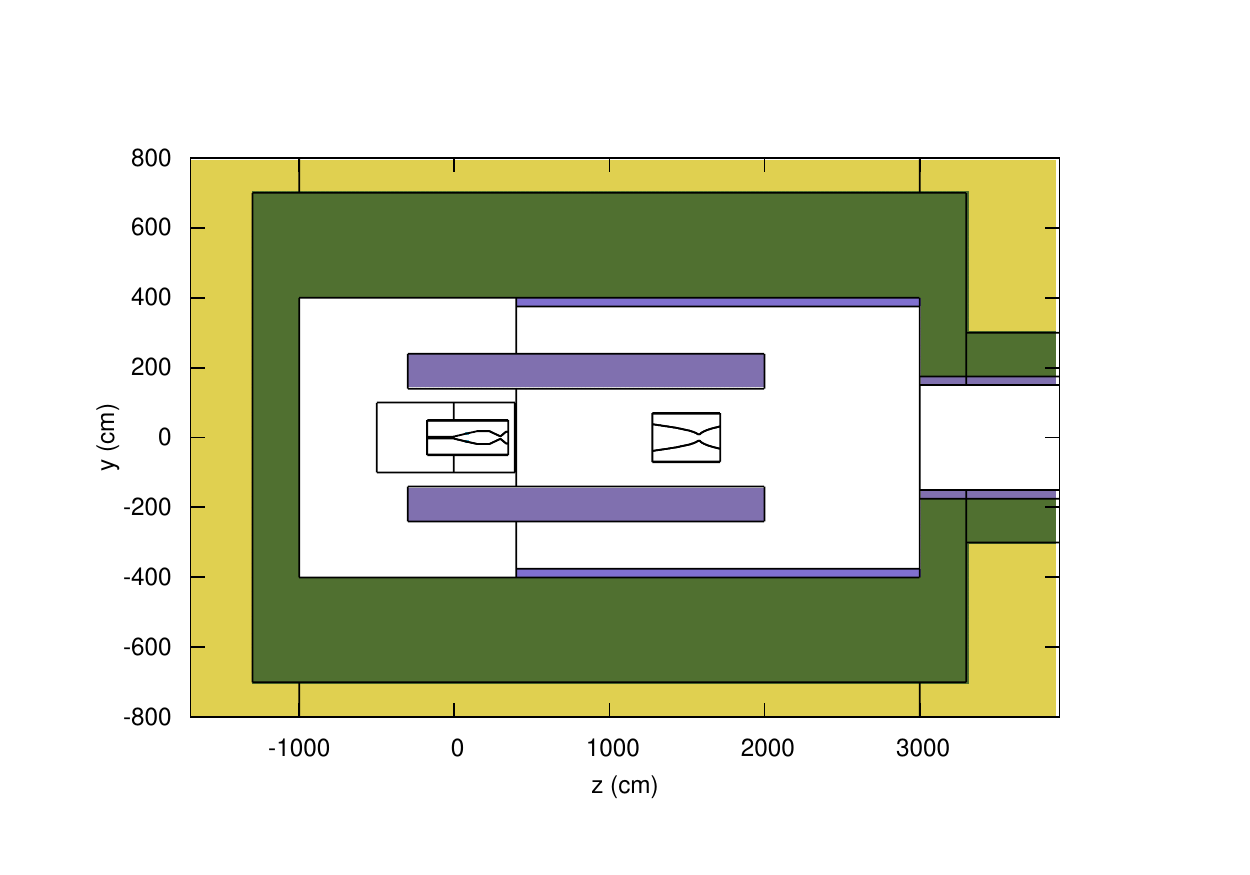}
\includegraphics[width=\the\figwidth]{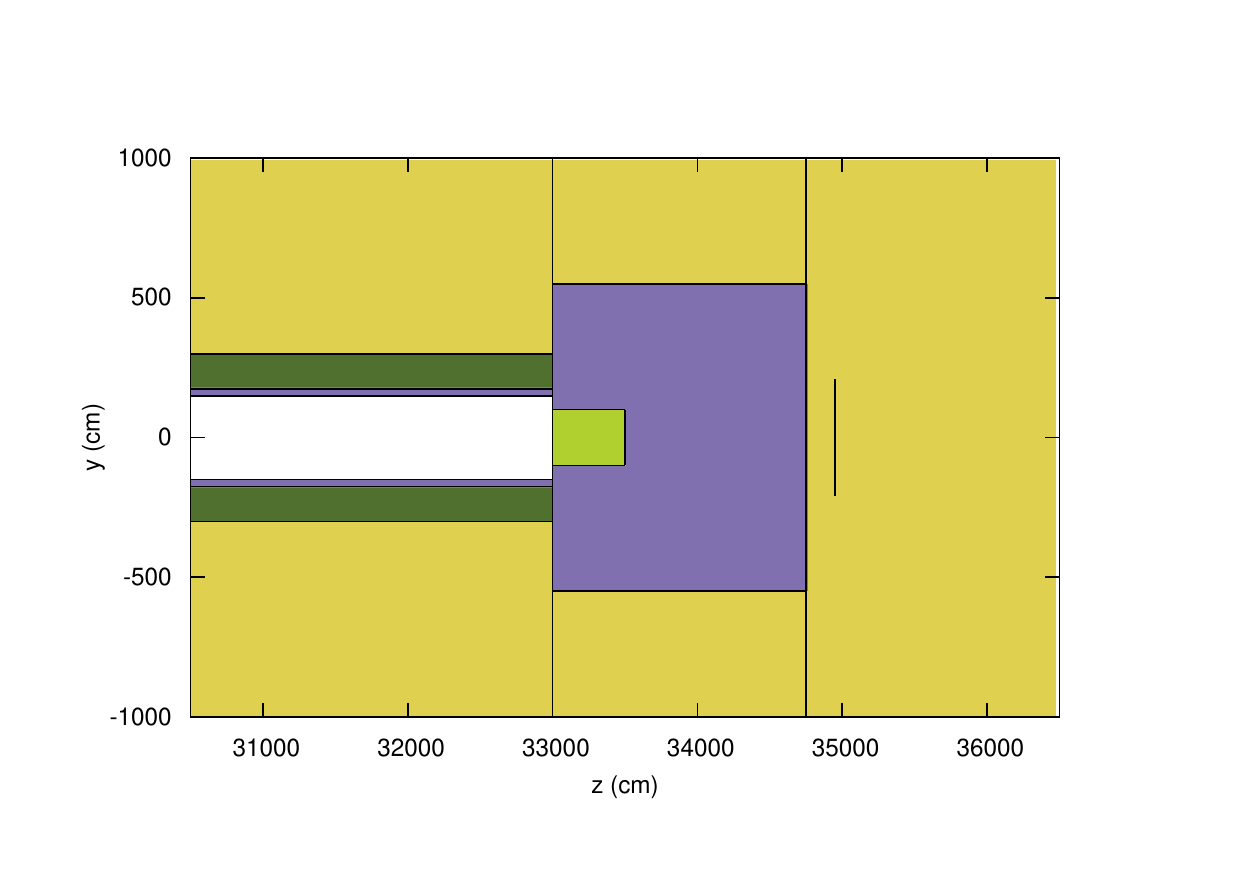}
\caption{FLUKA simulation of the CN2PY neutrino beamline. The target and both focusing horns are shown on the left and the end of the decay volume with the beam dump is shown on the right.}
\label{fig:beamlayout}
\end{figure}

A model of the neutrino beamline has been developed for the neutrino flux calculation using the latest version of the FLUKA \cite{Ferrari:2005zk} package. In the model, the hadron production target is a solid graphite cylinder. The focusing optics consists of two aluminum horns. The target is fully inserted into the first horn in order to maximize the collection of the low energy pions which contribute to the neutrino flux below 2 GeV (around the 2nd oscillation maximum). 
%At this stage no support system for the target and horns has been modeled and the components are simply placed in an empty environment representing the target station hall. 
The decay tunnel is located 30 m downstream of the target and is modelled as 300 m long and 3 m in diameter cylinder. A hadron beam stop (beam dump) is placed at the end of the decay volume. The 2D model layout of the neutrino beamline is shown in Figure~\ref{fig:beamlayout}.

\begin{figure}[t]
\center
\includegraphics[width=0.8\textwidth]{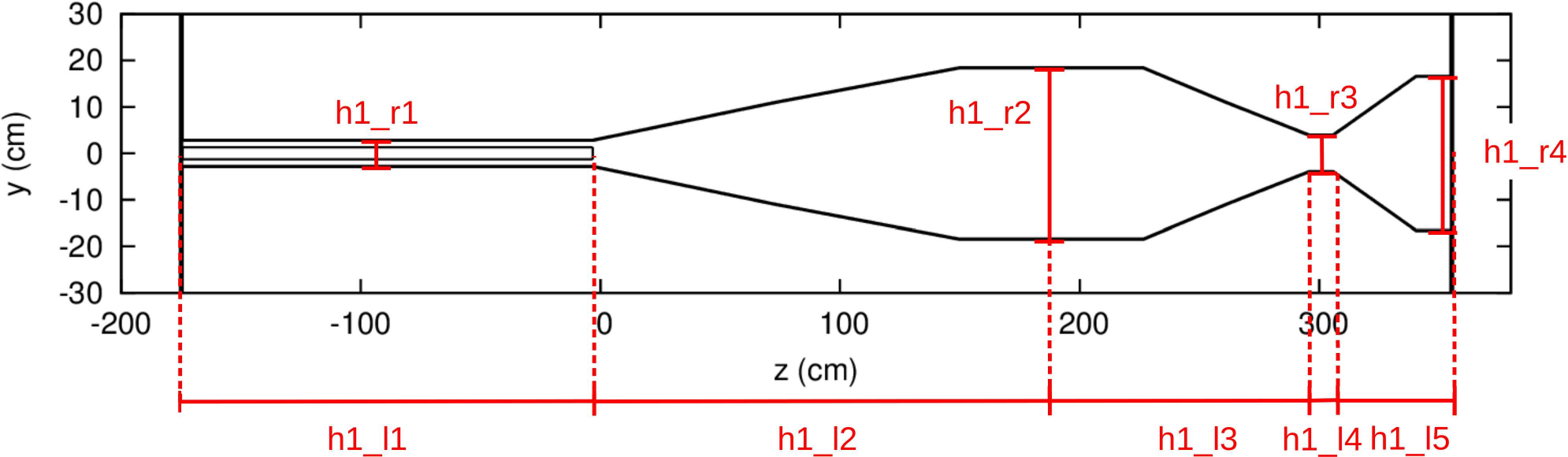}
\includegraphics[width=0.8\textwidth]{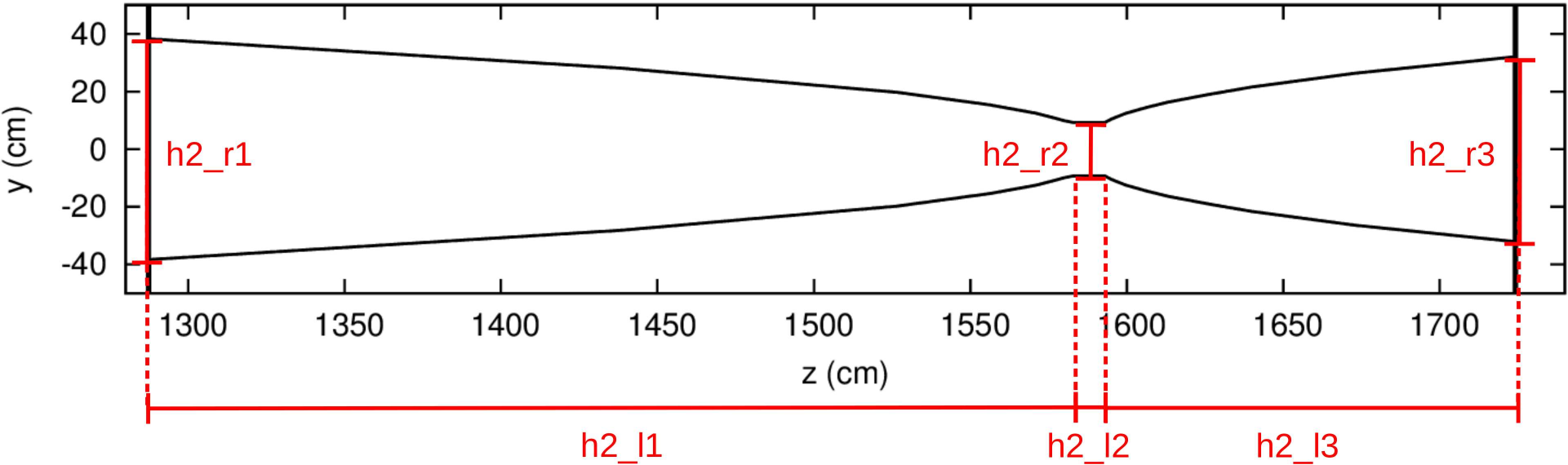}
\caption{Design layout of the 1st horn and target (top) and 2nd horn (bottom).}
\label{fig:beam_par_description}
\end{figure}

The geometry of the 1st horn adopted in this study differs significantly from the simple parabolic horn that was
assumed in the LBNO EoI~\cite{Stahl:2012exa}. The new design aims at improving the collection of low energy secondaries that exit the target at steep angles above 100~mrad, which could not be efficiently bent in the original design. The geometry of the 1st horn, shown in Figure~\ref{fig:beam_par_description}, is described in terms of ten parameters that define the structure of its inner conductor and the overall dimensions (See Table~\ref{tab:opticsparam}). The upstream part of the horn functions as a collector, which aims at minimising the divergence of the secondaries. The downstream part, with its elliptical inner conductor shape, attempts to focus particles along the beam axis. The 2nd horn or reflector, also shown in Figure~\ref{fig:beam_par_description}, has a more typical inner conductor shape that consists of two ellipsoidal sections. Its geometry is described in terms of seven parameters that define the shape of the inner conductor and the horn dimensions. 

\begin{table}[ht]
\centering
\begin{tabular}{l c c|c}
\hline
\hline
Parameter & Symbol & Unit & Parameter range \\
\hline
target radius & $r_{tgt}$ & cm & 0.4-1.5 \\
target length & $l_{tgt}$ & cm & 80-200  \\
\hline
circulating current in horn & $I_{H}$ & kA & 150-300      \\
circulating current in reflector & $I_{R}$ & kA & 150-250 \\
\hline
distance horn-reflector & $d_{HR}$ & m & 1-20 \\
\hline
horn length 1st part & h1\_l1 & cm  & fixed to $l_{tgt}$ \\
horn length 2nd part & h1\_l2 & cm  & 125-208  \\
horn length 3st part & h1\_l3 & cm  & 78-140   \\
horn length 4st part & h1\_l4 & cm  & 0-10     \\
horn length 5st part & h1\_l5 & cm  & 25-70    \\
horn 1st inner radius & h1\_r1 & cm & fixed to $r_{tgt}$+0.6 \\
horn 2nd inner radius & h1\_r2 & cm & 7-40    \\
horn 3rd inner radius & h1\_r3 & cm & 2.5-16  \\
horn 4th inner radius & h1\_r4 & cm & 2-20    \\
horn outer radius & h1\_r & cm & fixed to h1\_r2+30    \\
\hline
reflector length 1st part & h2\_l1 & cm & 50-300 \\
reflector length 2nd part & h2\_l2 & cm & 3-20   \\
reflector length 3rd part & h2\_l3 & cm & 50-300 \\
reflector 1st inner radius & h2\_r1 & cm & 10-40 \\
reflector 2nd inner radius & h2\_r2 & cm & 2-10  \\
reflector 3rd inner radius & h2\_r3 & cm & 10-40 \\
reflector outer radius & h2\_r & cm & fixed to h2\_r1+30  \\
\hline \hline
\end{tabular}
\caption{List of parameters describing beam optics and their respective allowed value ranges.}
\label{tab:opticsparam} 
\end{table}

\subsection{Flux optimisation}

The parameters to be optimised are listed in Table~\ref{tab:opticsparam}. These encapsulate the essential information about the target and horns geometries, the spatial orientation of components, and the focusing strength of the horns. Each parameter is allowed to vary independently from other parameters within a pre-set range, which is also shown Table~\ref{tab:opticsparam}. We use genetic algorithm, implemented in the DEAP toolkit \cite{DEAP_JMLR2012}, to find the optimal values for the parameters. The genetic algorithm is a search heuristic that mimics the process of natural selection in order to generate useful solutions to optimisation and search problems. A population of candidate solutions called individuals, in our case different beamline configurations, is evolved towards a better solution. Each candidate solution has a set of properties, called chromosomes, which in this case are the parameters in Table~\ref{tab:opticsparam}. For each generation, the fitness of each individual is evaluated and the best performing individuals are selected to randomly recombine and mutate their chromosomes to produce new individuals. This iterative process is carried out until no further improvement in fitness parameter is observed in the population. 
At this stage, we include engineering requirements on the optimisation process by fixing reasonable physical ranges of parameters. The primary goal is to find a configuration that delivers the optimal performance in terms of experimental sensitivity to $\delta_{CP}$. At the later stage of the implementation project, detailed design will refine these constraints taking into account
engineering constraints.

\begin{figure}[t]
\center
\includegraphics[width=0.95\textwidth]{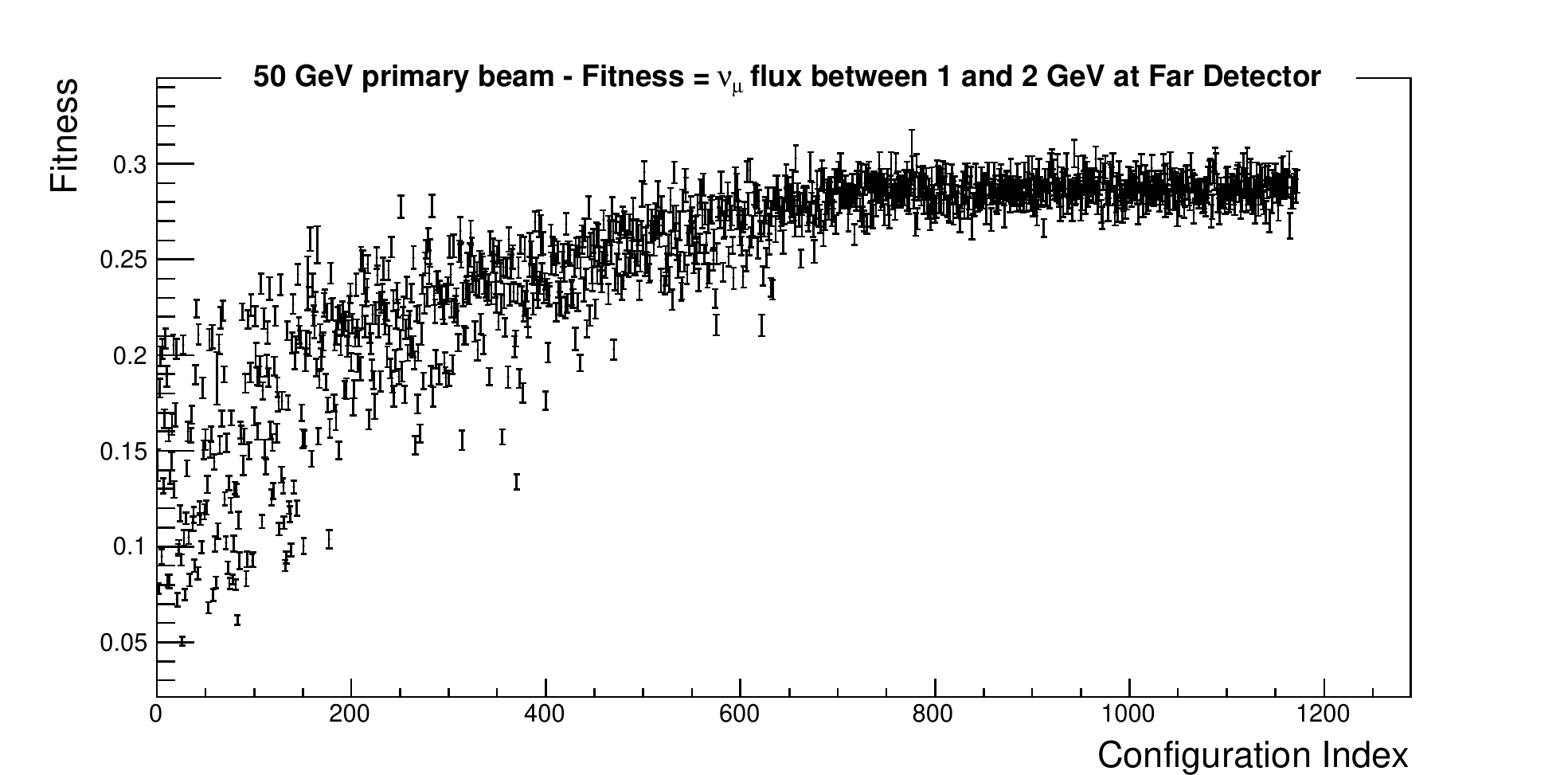}
\caption{Evolution of the fitness parameter during LE optimisation of the HPPS-based neutrino beam. The horizontal
axis represents
the chronological index of the beamline configurations generated by the algorithm.}
\label{fig:lefitnessevol}
\end{figure}

A critical part of the optimisation procedure is a suitable definition of the fitness criteria which is the quantity that the algorithm will attempt to maximize. To explore different possible energy windows for the neutrino beam we chose the following three different fitness criteria:
\begin{itemize}
\item {\bf High Energy optimisation (HE)}: maximization of the integral of $\nu_{\mu}$ flux in a 0-6 GeV energy window. In this case, the optimisation should generate beam optics configurations producing wide band beams covering both first and second oscillation maxima.
\item {\bf Low Energy optimisation (LE)}: maximization of the integral of $\nu_{\mu}$ flux in a 1-2 GeV energy window. In this case, beams optics configurations generating a neutrino flux mainly at lower energies around the second maximum oscillation maximum should be obtained.
\item {\bf CPV based optimisation using GloBeS (GLB)}: maximization of total $\delta_{\textrm{CP}}$ sensitivity as computed by GloBeS \cite{Huber:2004ka} with all the systematics uncertainties turned off. 
\end{itemize}

As an example, Figure~\ref{fig:lefitnessevol} shows the evolution of the fitness parameter for the LE optimisation of the HPPS-based neutrino beam. Typically, when the algorithm converges, it gives an ensemble of configurations each considered to be optimal from the point of view of the fitness criterion. One has to then choose the best candidate based on considerations related the engineering implementation for a given option.

\begin{figure}[t]
\begin{center}
\includegraphics[width=\the\figwidth]{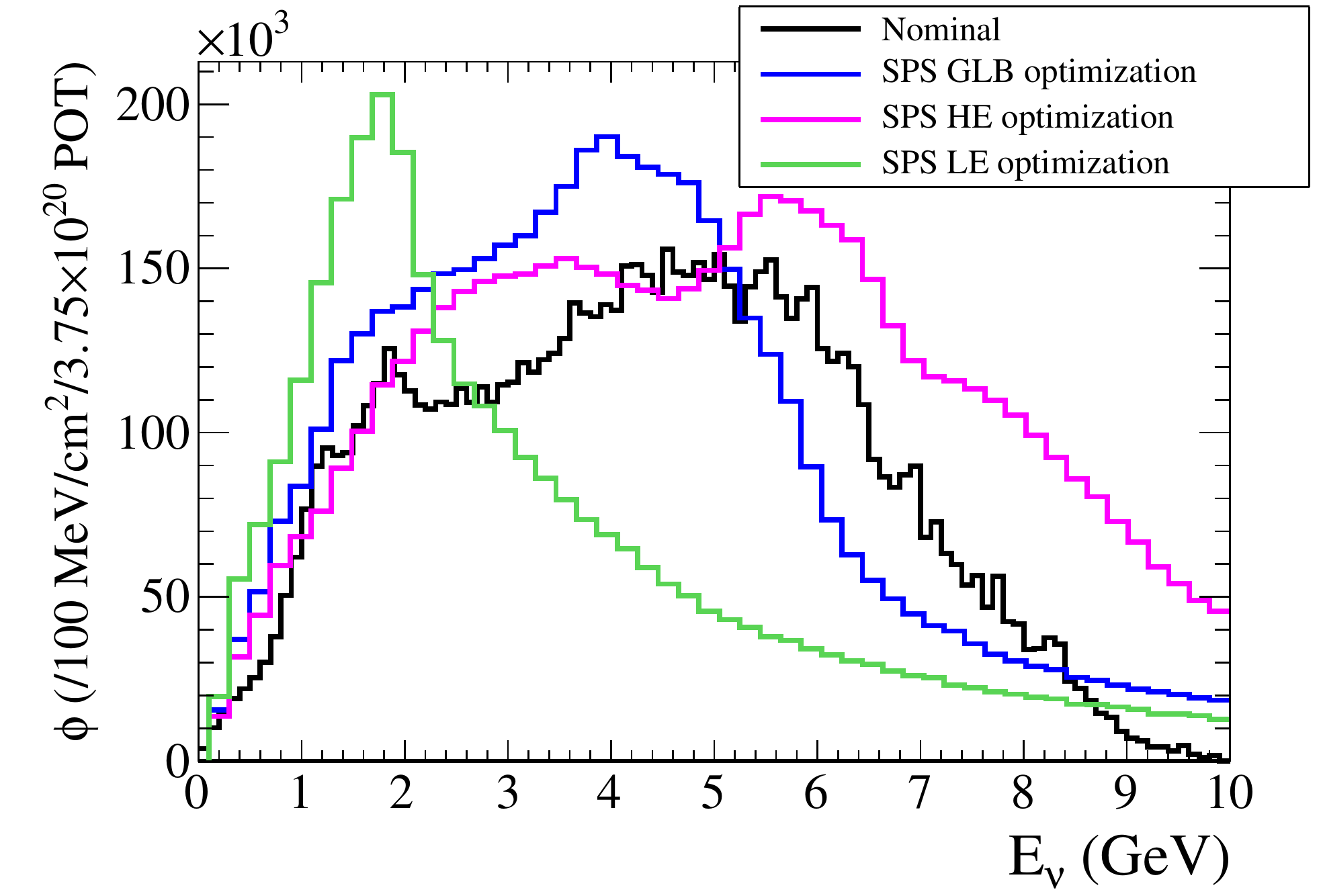}
\includegraphics[width=\the\figwidth]{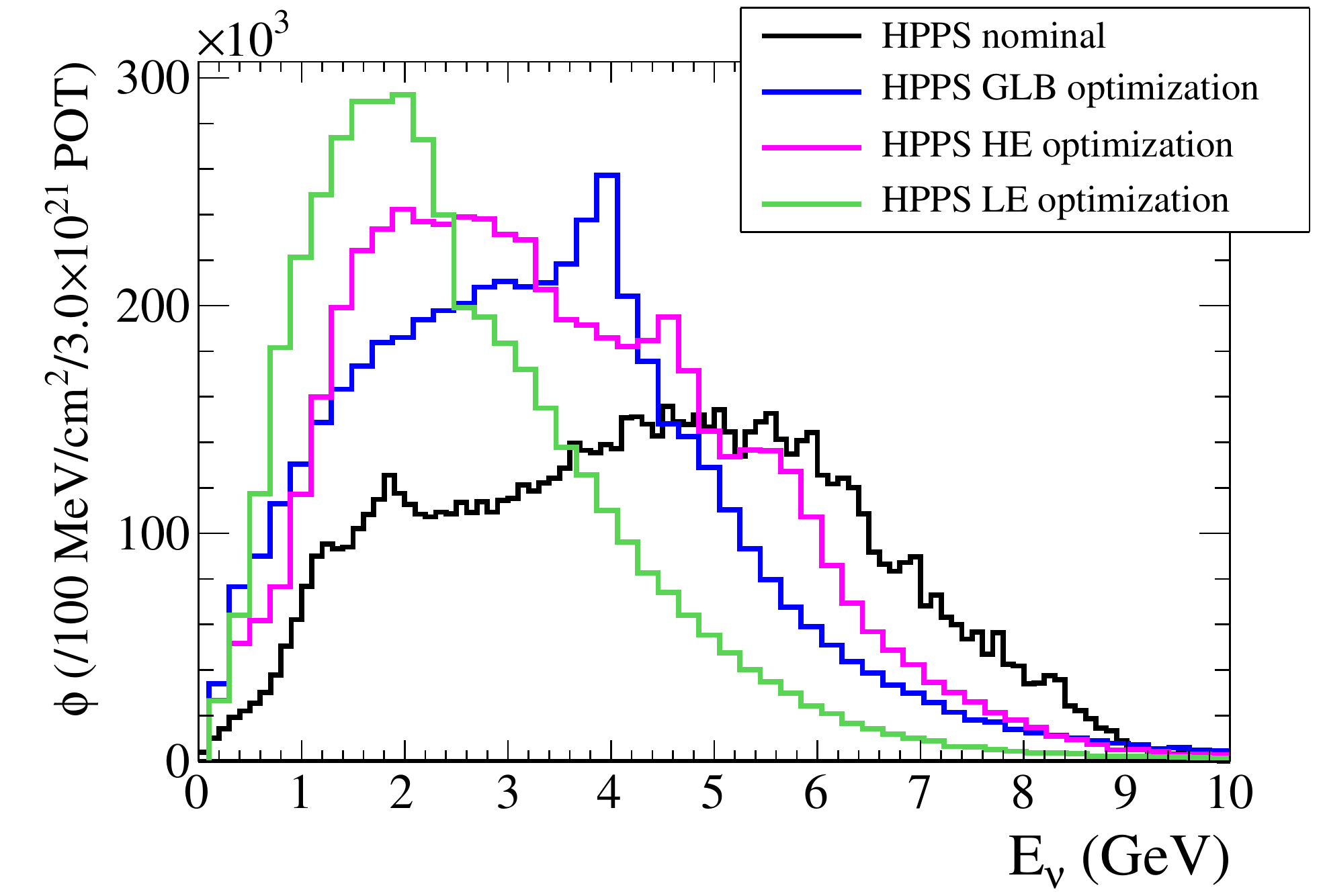}
\caption{Energy spectra of neutrino fluxes for different optimisations for SPS (left) and HPPS (right) proton beam options. The nominal LBNO flux from \cite{Stahl:2012exa} is also shown.}
\label{fig:cern_beams_optim}
\end{center}
\end{figure}

\begin{figure}[t]
\begin{center}
\includegraphics[width=\the\figwidth]{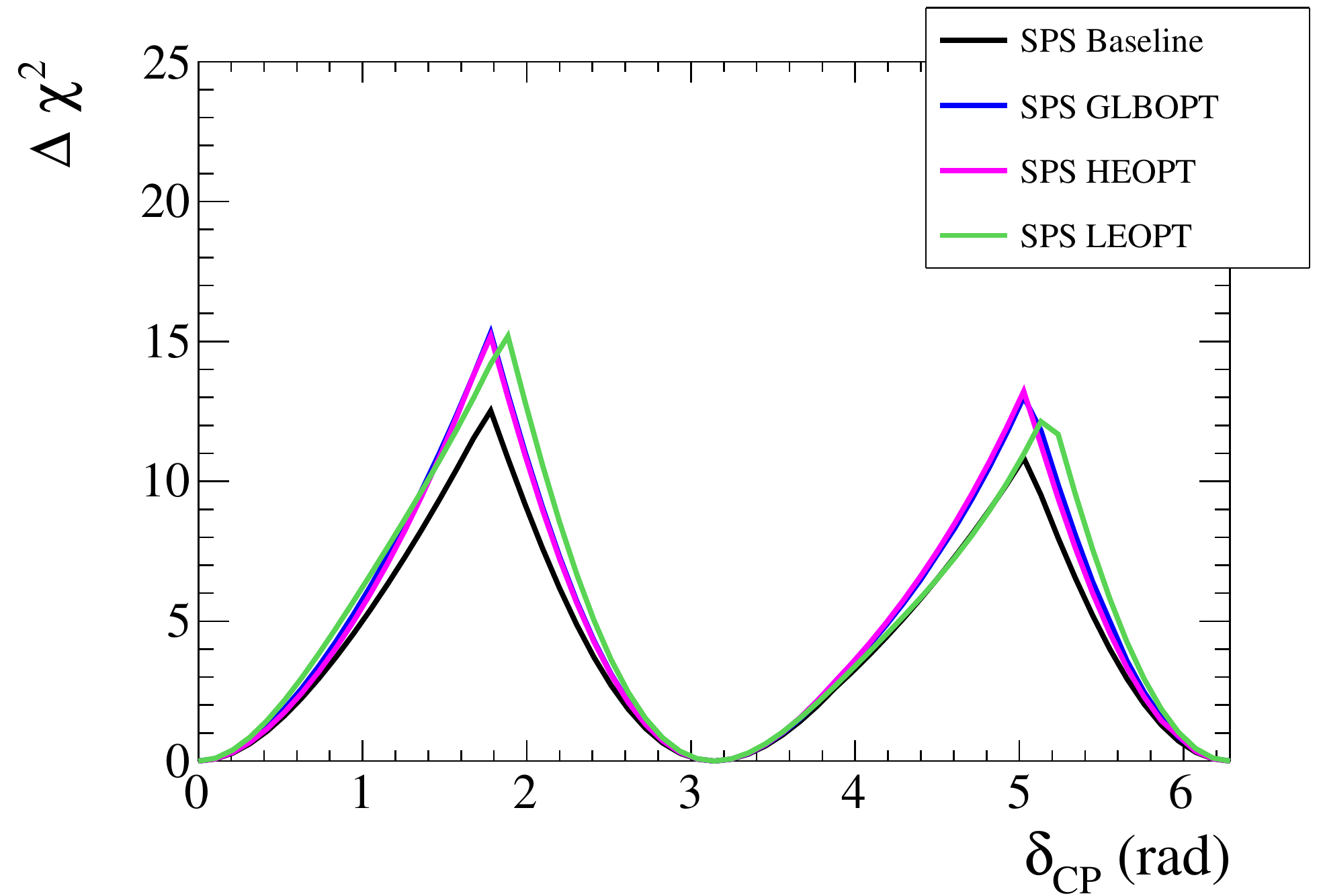}
\includegraphics[width=\the\figwidth]{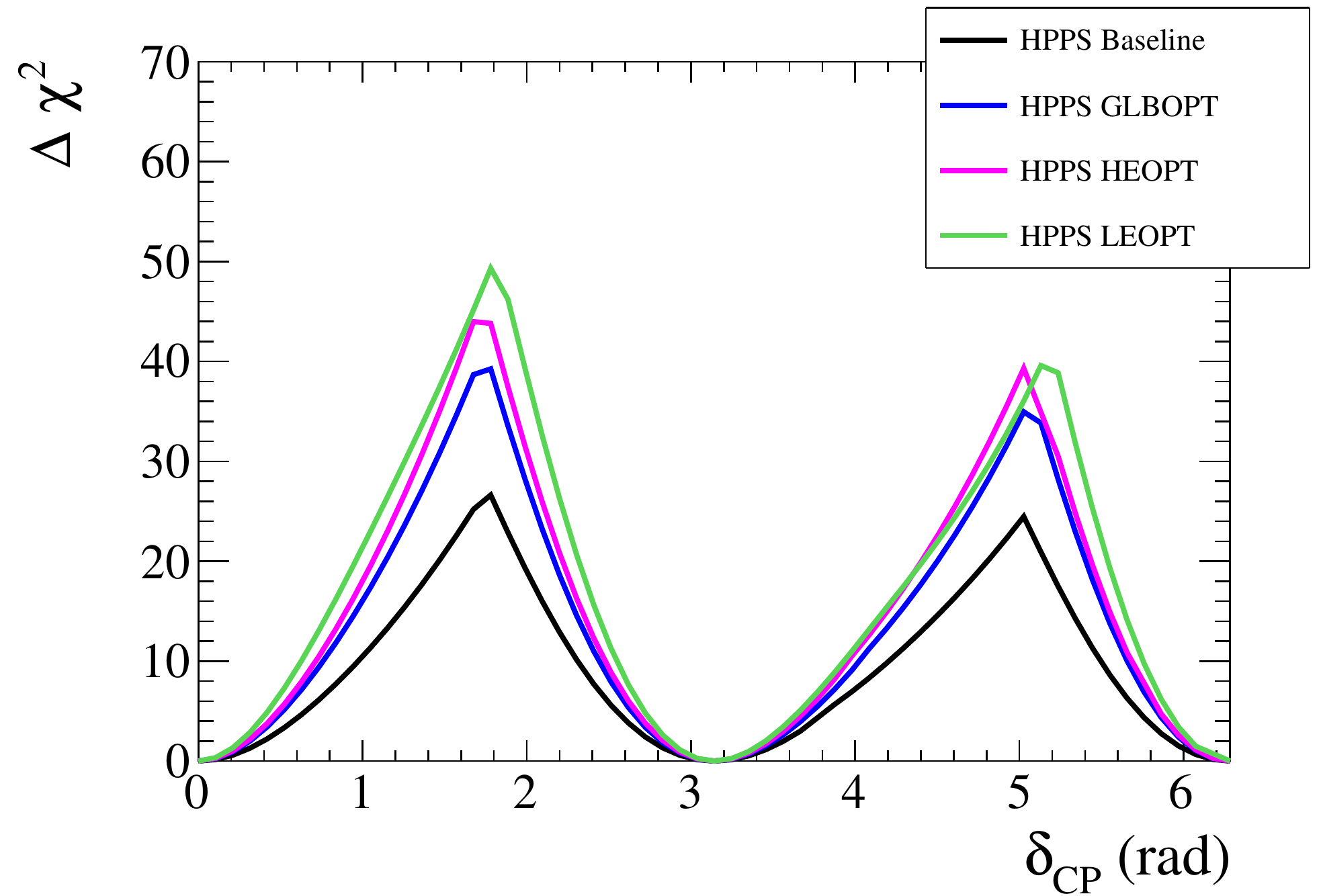}
\caption{Comparison of expected CPV sensitivity for different beam optimisations for SPS (left) and HPPS (right) proton beam options and LBNO20 detector configuration. A total exposure of $15\times 10^{20}$ ($30\times 10^{21}$) POT is taken for SPS (HPPS) beam with 75\% of the running time devoted to the running in the neutrino mode. The value of $\sin^2{\theta_{23}} = 0.5$ is assumed.}
\label{fig:cern_beams_optim_cpv}
\end{center}
\end{figure}

\begin{table} [h]
\centering
\begin{tabular}{c|c|c|c|c|c|c}
\hline
\hline
       & \multicolumn{3}{|c|}{SPS beam} & \multicolumn{3}{|c}{HPPS beam}  \\ 
       & Peak I & Peak II & $+F_{3\sigma}$ &  Peak I & Peak II & $+F_{5\sigma}$ \\
\hline
Base   & 12.5   & 10.8    &              &  26.6   &  24.5   &              \\
GLBOPT & 15.3   & 13.0    &  +8\%        &  39.2   &  35.0   &  +19\%       \\
LEOPT  & 15.1   & 13.2    &  +8\%        &  49.3   &  40.0   &  +28\%       \\
HEOPT  & 15.3   & 13.2    &  +8\%        &  44.0   &  39.3   &  +24\%       \\
\hline
\hline
\end{tabular}
\caption{Comparison of different beam optimisations in terms of the maximum sensitivity in first (second) half of $\delta_{CP}$ plane referred to as ``Peak I'' (``Peak II'') and the improvement in coverage at a given confidence level -- $3\sigma$ ($5\sigma$) for SPS (HPPS) -- relative to the baseline beam option. See Figure~\ref{fig:cern_beams_optim_cpv} for the respective sensitivity curves.}
\label{tab:bmoptimcompare}
\end{table}

\begin{figure}[ht]
\center
\includegraphics[width=\the\figwidth]{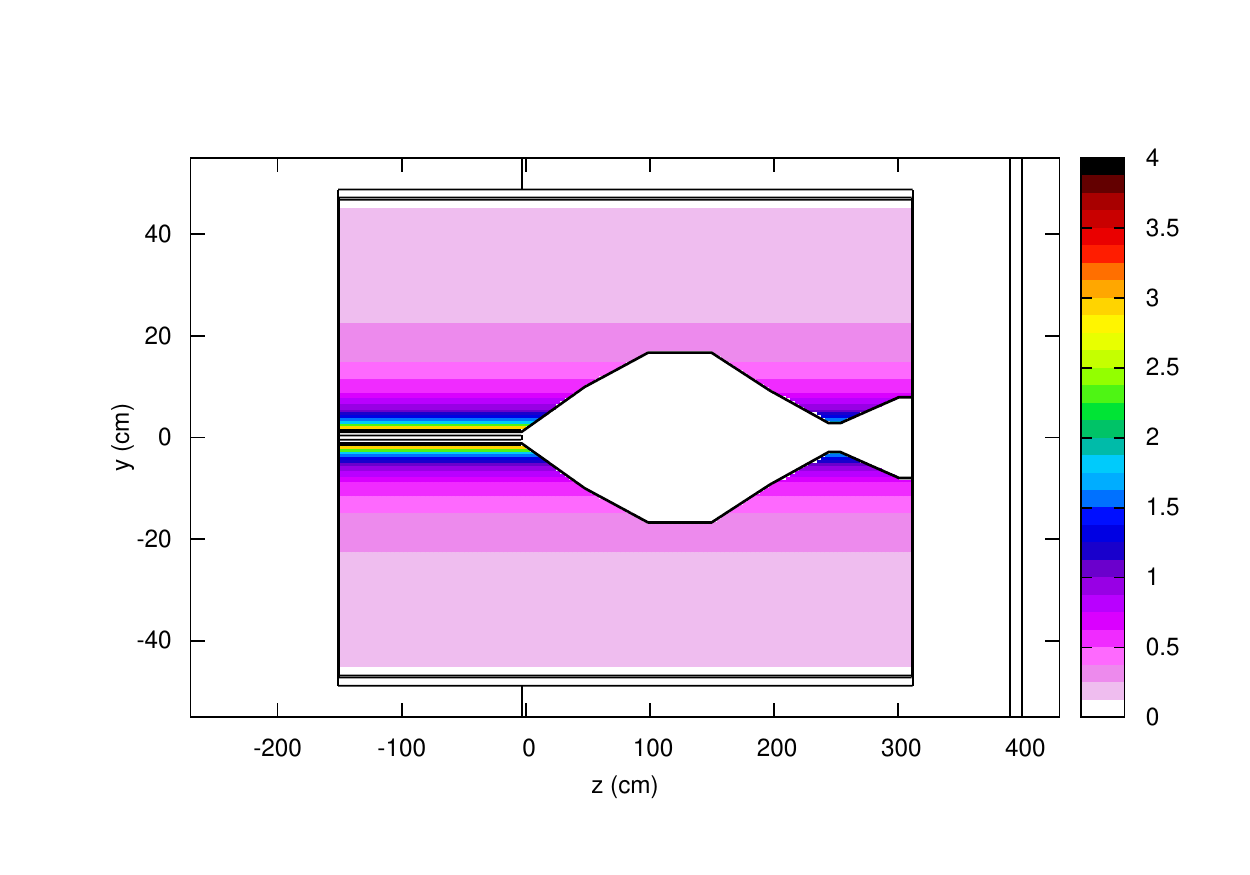}
\includegraphics[width=\the\figwidth]{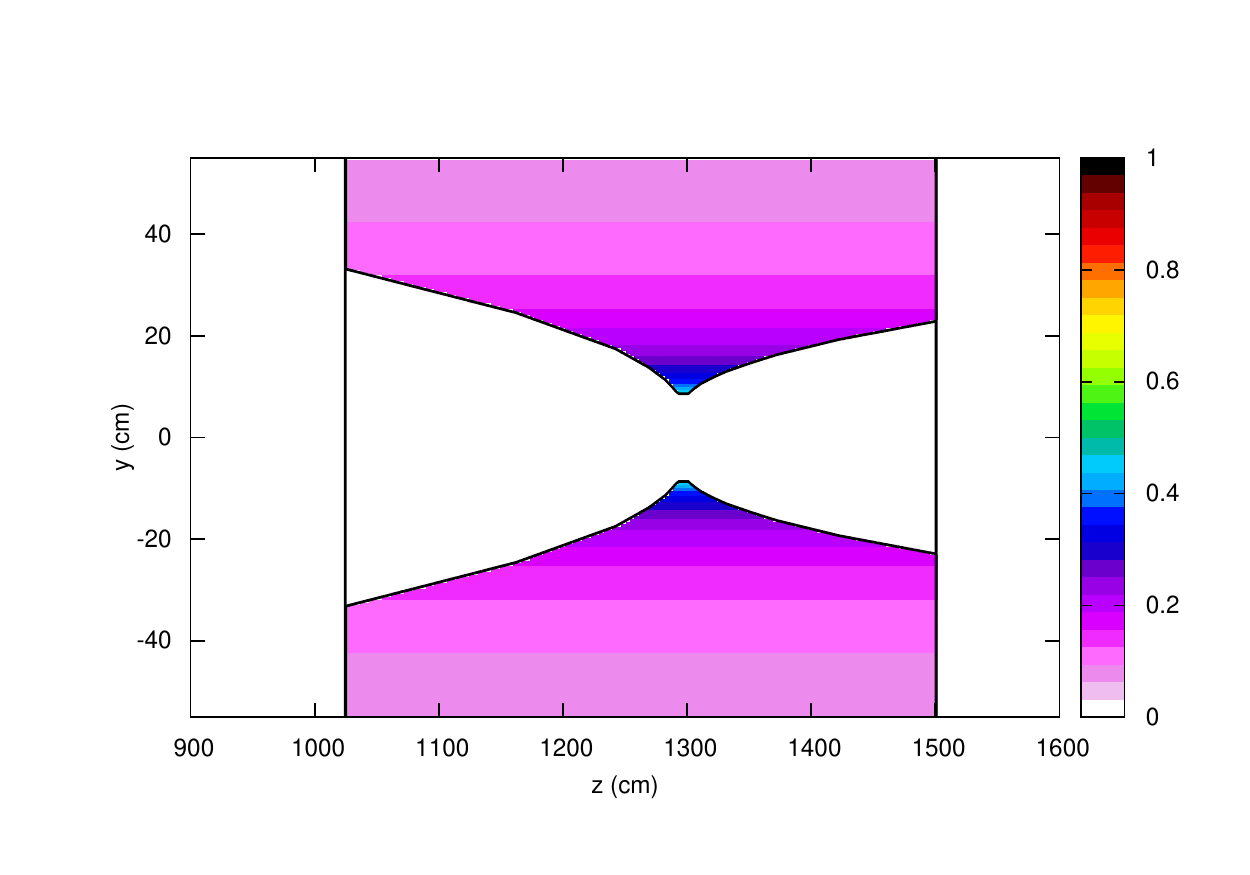}
\includegraphics[width=\the\figwidth]{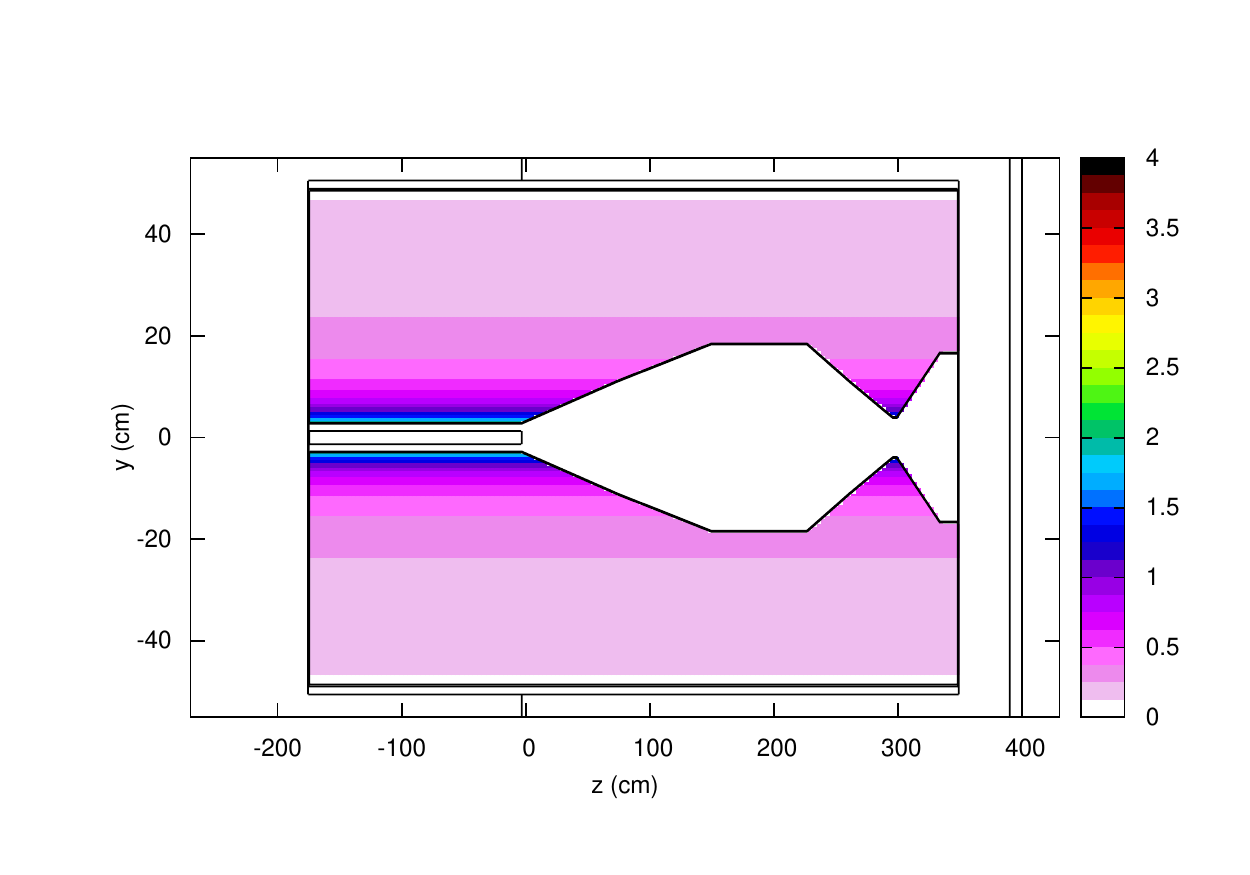}
\includegraphics[width=\the\figwidth]{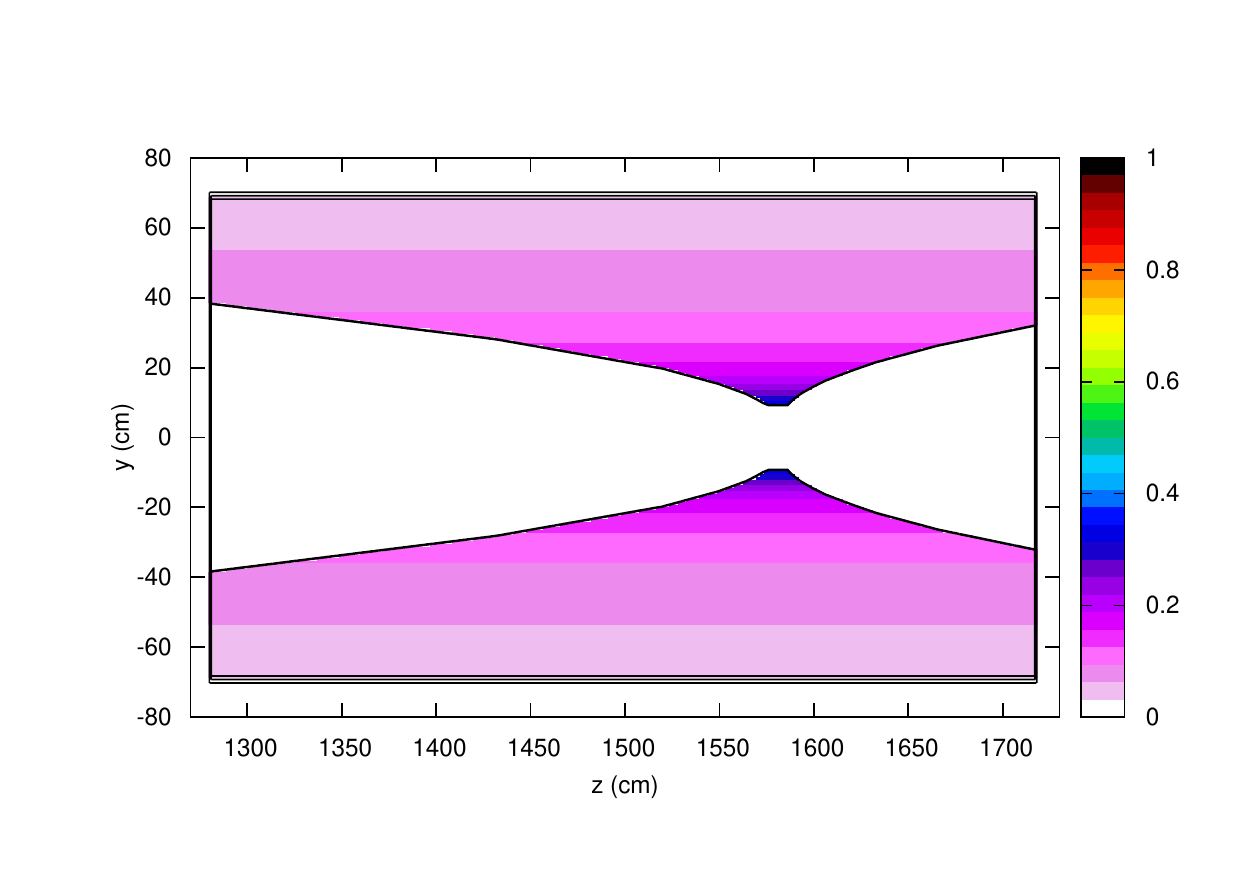}
\caption{View of the horn and target (left) and reflector (right) corresponding to the selected SPS GLB (panels above) and HPPS LE (panels below) configurations. The magnetic field strengths correspond to circulating currents of 281~kA (289~kA) and 198~kA (187~kA) for the 1st horn and 2nd horn respectively and SPS (HPPS) beam option. }
\label{fig:optimbeam_layout}
\end{figure}

The energy spectra of the $\nu_\mu$ flux for the best performing configurations in the three optimisation schemes are shown in Figure~\ref{fig:cern_beams_optim} for SPS and HPPS options. As can be seen in the figure, different optimisation schemes result in different neutrino energy spectra. To choose the best beam for the measurement of $\delta_{CP}$, we process each optimisation result through the full LBNO analysis framework (briefly described in Section~\ref{sec:analysis}) and calculate the sensitivity to CPV. 
The results will be discussed in Sections~\ref{sec:beamandcpv} and \ref{sec:results}, and are summarised in Table~\ref{tab:bmoptimcompare}. 
In general, each optimised beam offers a better performance over the nominal LBNO flux with a particularly large improvement achieved by the LE optimisation of the HPPS beam. 
In the case of the SPS optimisation, there is a small difference between the CPV sensitivities obtained with the SPS GLB, HE and LE.
%The LE optimisation has a significant flux in the lower energy region, however, the flux in the higher energy region is significantly
%suppressed compared to the GLB and HE. In this case, both low (around 2nd maximum) and high (around 1st) options yield similar
%sensitivities. For the HPPS, the lower primary proton energy compared to the SPS, allows an optimisation where the flux at low energy
%is significantly improved compared to the flux at higher energies. In this case, the HPPS LE is better than the HPPS GLB and HE.
In general, the fluxes obtained with the 50~GeV primary proton energy at the HPPS
are more favourable for CPV, compared to the 400~GeV of the SPS.
For the physics studies shown in the subsequent sections,
we chose for definiteness the SPS GLB and HPPS LE optimisations. 
The corresponding target-horn layouts for these configurations are shown in Figure~\ref{fig:optimbeam_layout} for 
both primary proton beam options.

%%%%%%%%%%%%%%%%%%%%%%%%%%%%%%%%%%%%%%%%%%%%%%%%%%%%%%%%%%%%%%%%%%%%%
%             Review of analysis method
%%%%%%%%%%%%%%%%%%%%%%%%%%%%%%%%%%%%%%%%%%%%%%%%%%%%%%%%%%%%%%%%%%%%%
\section{Analysis procedure}
\label{sec:analysis}
A primary source of information for the measurement of MH and CPV is the sample of electron-like (e-like) events. The signal consists of $\nu_e$ charge-current (CC) events with an electron neutrino from $\nu_\mu\rightarrow \nu_e$ oscillations. We consider four background contributions:
\begin{itemize}
\item{Intrinsic $\nu_e$ ($\bar{\nu}_e$) contamination in the beam (intrinsic $\nu_e$),}
\item{Electron events from $\nu_\tau$ CC interactions 
with subsequent leptonic $\tau$ decay ($\nu_\tau\rightarrow \tau \rightarrow e\nu\nu$ contamination),}
\item{Events with a $\pi^0$ from neutral current $\nu_\mu$ interaction mis-identified as an electron (NC $\pi^0$),}
\item{Mis-identified muons from $\nu_\mu$ CC interactions (mis-id $\nu_\mu$).}
\end{itemize}
In order to better differentiate signal events from background we use both the reconstructed neutrino energy $E_\nu^{rec}$ and the missing transverse momentum $p_T^{miss}$ as observables~\cite{Stahl:2012exa,Agarwalla::2013kaa}. 
Mathematically the number of the e-like events in a given $E_\nu^{rec}-p_T^{miss}$ bin is:
\begin{equation}
\begin{split}
n_{e}(E_\nu^{rec},p_T^{miss}; \mathbf{o}, \mathbf{f}) &= f_{sig}n_{e-sig}(E_\nu^{rec},p_T^{miss};\mathbf{o}) \\ &+ f_{\nu_e}n_{\nu_e}(E_\nu^{rec},p_T^{miss};\mathbf{o}) + f_{\nu_\tau} n_{e,\nu_\tau}(E_\nu^{rec},p_T^{miss};\mathbf{o}) \\ &+ f_{NC}(n_{NC\pi^0}(E_\nu^{rec},p_T^{miss};\mathbf{o})+n_{mis-\nu_\mu}(E_\nu^{rec},p_T^{miss};\mathbf{o})),
\end{split}
\label{eq:nelike}
\end{equation}
where $n_{e-sig}$, $n_{\nu_e}$, $n_{e,\nu_\tau}$, $n_{NC\pi^0}$, and $n_{mis-\nu_\mu}$ are the number of events for signal, intrinsic beam $\nu_e$, electrons from tau decay, neutral current, and mis-identified $\nu_\mu$, respectively. The event rate also depends on the oscillation parameters $\mathbf{o}$ and systematic parameters $\mathbf{f}$ which parametrize the normalization uncertainties in signal and background contributions.

In addition to the e-like sample  we also look at a sample of $\mu$-like events which come primarily from $\nu_\mu$ CC interactions in the disappearance channel ($\nu_\mu \rightarrow \nu_\mu$ survival probability) with a small contribution from muon events originating from $\tau \rightarrow \mu$ decays. This allows us to constrain the atmospheric oscillation parameters: $|\Delta m^2_{31}|$ and $\sin^2{\theta_{23}}$. The number of $\mu$-like events in a given bin of reconstructed neutrino energy is the sum of signal ($n_{\mu-sig}$) and $\tau\rightarrow \mu$ background, $n_{\mu,\nu_\tau}$, contributions:
\begin{equation}
n_{\mu}(E_\nu^{rec}; \mathbf{o}, \mathbf{f}) = f_{sig} n_{\mu-sig}(E_\nu^{rec};\mathbf{o}) + f_{\nu_\tau} n_{\mu,\nu_\tau}(E_\nu^{rec};\mathbf{o}).
\label{eq:nmulike}
\end{equation}

In Table~\ref{tab:oscparam_may2014}, we summarize the values of the parameters that are used to define oscillation probabilities. The parameters which are not labeled as ``exact'' are allowed to vary within their prior error values.
\begin{table}[h]
\centering
\begin{tabular}{c|c|c}
\hline
\hline
Parameter & Central value & Fractional uncertainty \\
\hline
Baseline $L$ (km) & 2300 & exact \\
$\sin^2{\theta_{12}}$  & 0.306  & exact  \\
$\sin^2{2\theta_{13}}$ & 0.09   & 3\%    \\
$\sin^2{\theta_{23}}$  & 0.45-0.55 & 5\% \\
$|\Delta m^2_{31}| \times 10^{-3} (\mbox{eV}^2)$ & 2.50 & 2\%   \\
$\Delta m^2_{21} \times 10^{-5} (\mbox{eV}^2)$   & 7.45 & exact \\
Average matter density $\rho (\mbox{g/cm}^3)$ & 3.2 & 4\% \\
\hline
\hline
\end{tabular}
\caption{Summary of the parameter values used to calculate neutrino oscillation probabilities. The parameters whose values are not assumed to be exact are treated as nuisance parameters inside the analysis framework (allowed to vary during the fit).}
\label{tab:oscparam_may2014}
\end{table}

Presently the octant of $\theta_{23}$ is unknown. The most precise recent measurement of this mixing angle by the T2K experiment \cite{Abe:2014ugx} indicates maximal mixing: 
\begin{equation}
 \sin^2{\theta_{23}} = 0.511 \pm 0.055 
\end{equation}
In Table.~\ref{tab:oscparam_may2014}, we therefore indicate the value of $\sin^2{\theta_{23}}$ to be in the window between 0.45 and 0.55 (approximately $1\sigma$ range of the T2K measurement). 
The global analysis of all neutrino oscillation data \cite{GonzalezGarcia:2012sz} prefers a lower value of 0.45 for $\sin^2{\theta_{23}}$. In Section~\ref{sec:results}, we will briefly discuss what effect the unknown octant has on the sensitivity to CPV, but in general we will take the value of $\sin^2{\theta_{23}}$ to be 0.45 in our studies.

The assumptions on the normalization uncertainties are listed in Table~\ref{tab:normerr_may2014}. Similar to the oscillation parameters these systematics are included as nuisance parameters inside the fit framework. 
\begin{table}[h]
\centering
\begin{tabular}{c|c|c}
\hline
\hline
Parameter & Central value & Fractional uncertainty \\
\hline
Signal normalization ($f_{sig}$) & 1 & 3\% \\
Beam electron normalization ($f_{\nu_e}$) & 1 & 5\% \\
Tau event normalization ($f_{\nu_\tau}$) & 1 & 20\% \\
$\nu$ NC and $\nu_\mu$ CC background ($f_{NC}$) & 1 & 10\% \\
\hline
\hline
\end{tabular}
\caption{Normalization uncertainties on signal and background events.}
\label{tab:normerr_may2014}
\end{table}

We perform a joint fit with e-like and $\mu$-like event samples to evaluate the sensitivity of the experiment to the CPV and MH. In the fit the following $\chi^2$ 
\begin{equation}
\chi^2 = \chi^2_{appear} + \chi^2_{disa} + \chi^2_{syst},
\label{eq:totchi2}
\end{equation}
is minimized with respect to the oscillation parameters $\mathbf{o}$ that are not fixed and systematic parameters $\mathbf{f}$ (Tables~\ref{tab:oscparam_may2014} and~\ref{tab:normerr_may2014}).

The $\chi^2_{appear}$ is the term corresponding to the electron appearance and is given by: 
\begin{equation}
\begin{split}
\chi^2_{appear} = 2\sum_{+/-}\sum_{E_\nu^{rec},p_T^{miss}} & n_{e}(E_\nu^{rec},p_T^{miss}; \mathbf{o}_{test}, \mathbf{f}_{test}) - n_{e}(E_\nu^{rec},p_T^{miss}; \mathbf{o}_{true}, \mathbf{f}_{true}) \\
& + n_{e}(E_\nu^{rec},p_T^{miss}; \mathbf{o}_{true}, \mathbf{f}_{true})\ln{\frac{n_{e}(E_\nu^{rec},p_T^{miss}; \mathbf{o}_{true}, \mathbf{f}_{true})}{n_{e}(E_\nu^{rec},p_T^{miss}; \mathbf{o}_{test}, \mathbf{f}_{test})}},
\end{split}
\label{eq:chi2appear}
\end{equation}
where the subscript \textit{true} (\textit{test}) refers to the true (test) values of the $\mathbf{o}$ and $\mathbf{f}$ parameters. The \textit{true} parameters are those chosen by Nature, while \textit{test} refer to the parameter at which we compute the likelihood with respect to the \textit{true} value.
The first sum  in Eq.~\ref{eq:chi2appear} corresponds to adding the contributions from the neutrino and anti-neutrino beam running. 

The information from the disappearance channel is contained in the $\chi^{2}_{disa}$ term of total $\chi^{2}$ in Eq.~\ref{eq:totchi2}, which is given by
\begin{equation}
\begin{split}
\chi^2_{disa} = 2\sum_{+/-}\sum_{E_\nu^{rec}} & n_{\mu}(E_\nu^{rec}; \mathbf{o}_{test}, \mathbf{f}_{test}) - n_{\mu}(E_\nu^{rec}; \mathbf{o}_{true}, \mathbf{f}_{true}) \\
& + n_{\mu}(E_\nu^{rec}; \mathbf{o}_{true}, \mathbf{f}_{true})\ln{\frac{n_{\mu}(E_\nu^{rec}; \mathbf{o}_{true}, \mathbf{f}_{true})}{n_{\mu}(E_\nu^{rec}; \mathbf{o}_{test}, \mathbf{f}_{test})}}.
\end{split}
\label{eq:chi2disa}
\end{equation}

\begin{figure}[ht]
\begin{center}
\includegraphics[width=0.5\textwidth]{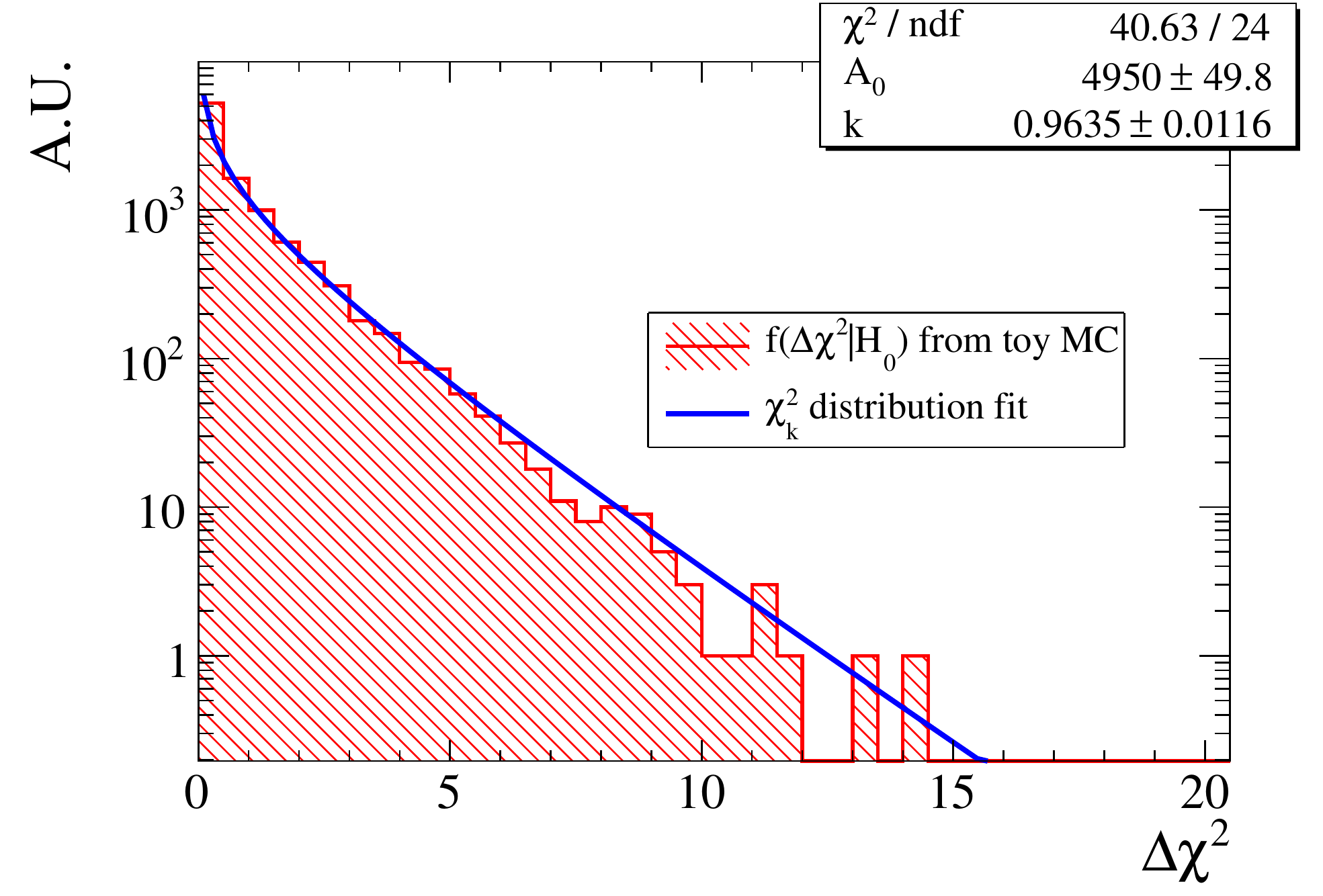}
\caption{Distribution of $\Delta\chi^2$ test statistic for $\delta_{CP} = 0$.}
\label{fig:DeltaChi2CPV}
\end{center}
\end{figure}

The last term in Eq~\ref{eq:totchi2}, $\chi^2_{syst}$, corresponds to the constraints applied on the oscillation and the systematic parameters. These constraints are assumed to be Gaussian symmetrical. Thus $\chi^2_{syst}$ is given by
\begin{equation}
\chi^2_{syst} = \sum_i \frac{(a_{0,i} - a_i)^2}{\sigma_{a_i}^2},
\end{equation}
where $a_{0,i}$ ($a_i$) is the prior (test) value of the \textit{i\textsuperscript{th}} parameter and $\sigma_{a_i}$ is the corresponding prior uncertainty.

The test statistic for CP violation is evaluated at a given value of $\delta_{CP}$ by minimizing the $\chi^2$ in Eq.~\ref{eq:totchi2} at the fixed test values of 0 and $\pi$. Thus one determines two quantities:
\begin{equation}
\begin{split}
&\Delta \chi^2_0 = \chi^2(\delta_{CP} = 0)   - \chi^2_{true} \\
&\Delta \chi^2_{\pi} = \chi^2(\delta_{CP} = \pi) - \chi^2_{true}
\end{split}
\end{equation}
and then takes
\begin{equation}
\label{eq:dchi2cpv}
\Delta \chi^2 = \min{(\Delta\chi^2_0, \Delta\chi^2_\pi)}.
\end{equation}

As discussed in \cite{Agarwalla::2013kaa} and again demonstrated here in Figure~\ref{fig:DeltaChi2CPV}, the distribution of $\Delta\chi^2$ in Eq.~\ref{eq:dchi2cpv} in the case CP is conserved (true value of $\delta_{CP}$ is either zero or $\pi$) follows a $\chi^{2}$ distribution with one degree of freedom. Consequently the median significance for CPV discovery (rejection of $\delta_{CP} \neq 0, \pi$ values by the experiment) can be evaluated as 
\begin{equation}
\label{eq:nsigdeltacp}
n\sigma = \sqrt{\Delta\bar{\chi}^2},
\end{equation}
where $\Delta\bar{\chi}^2$ is obtained from median value of the test statistic distribution under the CP non-conservation hypothesis. A more detailed discussion of the statistics for determining CPV (and also evaluating the sensitivity to MH) can be found in \cite{Agarwalla::2013kaa}.

%%%%%%%%%%%%%%%%%%%%%%%%%%%%%%%%%%%%%%%%%%%%%%%%%%%%%%%%%%%%%%%%%%%%%
%             Sensitivity to CPV
%%%%%%%%%%%%%%%%%%%%%%%%%%%%%%%%%%%%%%%%%%%%%%%%%%%%%%%%%%%%%%%%%%%%%
\section{Energy spectrum of neutrino beam and CPV sensitivity in LBNO}
\label{sec:beamandcpv}

In Figure~\ref{fig:cern_beams_optim_cpv}, we provide the  $\Delta \chi^2$ relevant for CPV (see Eq. \ref{eq:dchi2cpv}) as a function
of the true $\delta_{CP}$, obtained with different beam optimisations described in Section~\ref{sec:nuflux} and whose spectra are shown in Figure~\ref{fig:cern_beams_optim}. As already shown above, the optimised beams offer a better performance than over the baseline LBNO flux with a particularly large improvement achieved by the LE optimisation of the HPPS beam. For the results presented in the next sections,  we have chosen GLB and LE optimisations of SPS and HPPS primary beam,  respectively. In the following we will elucidate the correlation between the beam profiles of the different options, and the 
resulting sensitivity to CPV. 

\begin{figure}[t]
\begin{center}
\includegraphics[width=\the\figwidth]{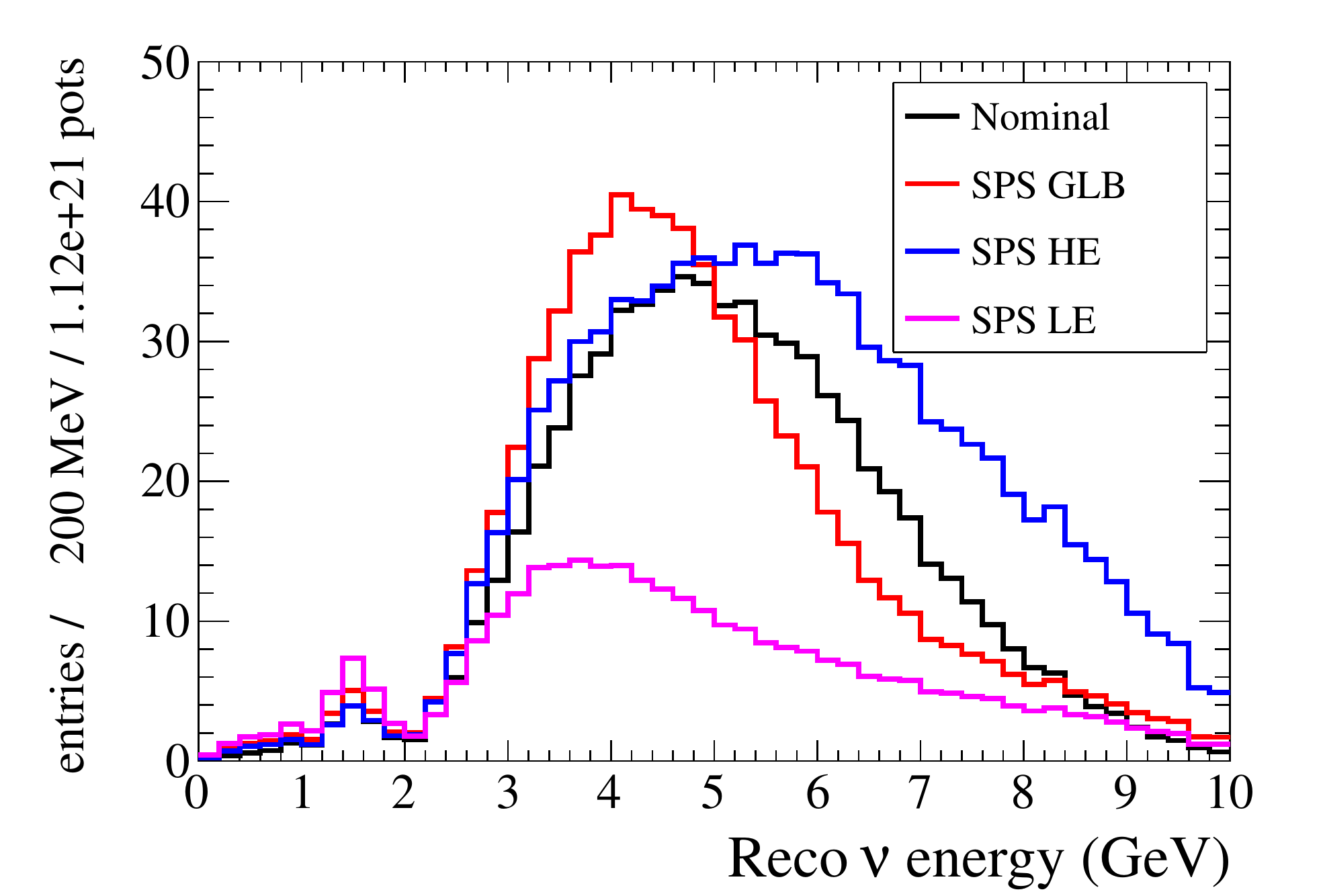}
\includegraphics[width=\the\figwidth]{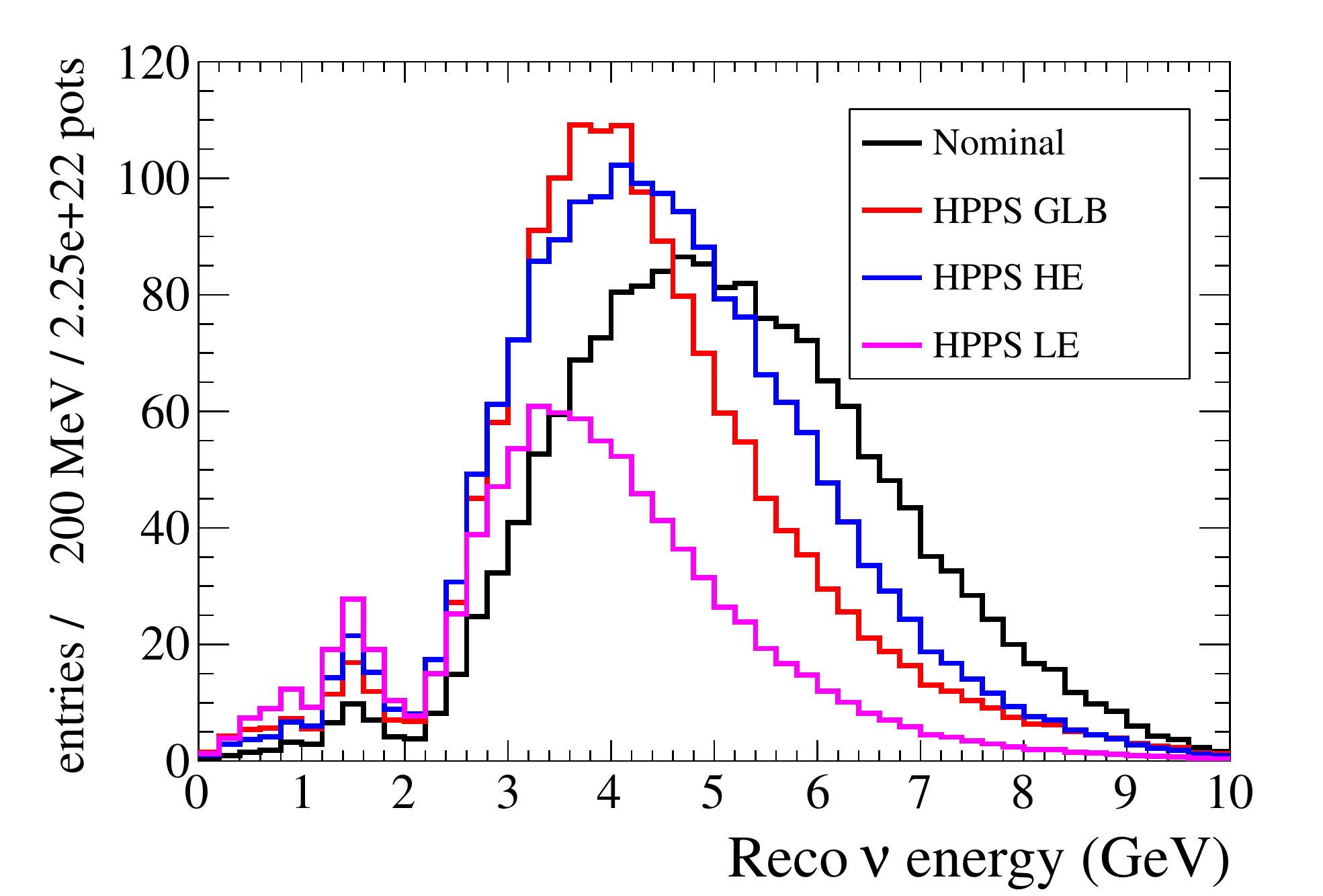}
\caption{Comparison of signal rates obtained with different neutrino beam optimisations for SPS (left) and HPPS (right) proton beams.}
\label{fig:compare_signue_optim}
\end{center}
\end{figure}

To qualitatively understand the results observed in Figure~\ref{fig:cern_beams_optim_cpv}, one needs to examine the expected event rates for each chosen beam optimisation. In Tables~\ref{tab:sps_elike_rates_cp0} and~\ref{tab:hpps_elike_rates_cp0}, we provide the breakdown of the expected contributions to the e-like event sample (both from signal and background) for SPS and HPPS beam options in the example of $\delta_{CP}=0$ and NH. In the case of expected signal rate, we also give the contributions at low energies (0--3 GeV) and high energies (3--10 GeV). In addition, the expected distributions of the reconstructed energy 
for the different beam optimisations are shown in Figure~\ref{fig:compare_signue_optim}.

%\begin{figure}[t]
%\begin{center}
%\includegraphics[width=\the\figwidth]{figures/cmproptim_medsense_cpv_sps_20kton.pdf}
%\includegraphics[width=\the\figwidth]{figures/cmproptim_medsense_cpv_hpps_20kton.pdf}
%\caption{Comparison of expected CPV sensitivity for different beam optimisations for SPS (left) and HPPS (right) proton beam options and LBNO20 detector configuration. A total exposure of $15\times 10^{20}$ ($30\times 10^{21}$) POT is taken for SPS (HPPS) beam with 75\% of the running time devoted to the running in the neutrino mode. The value of $\sin^2{\theta_{23}} = 0.5$ is assumed.}
%\label{fig:cern_beams_optim_cpv}
%\end{center}
%\end{figure}

\begin{table} [h]
\centering
\begin{tabular}{c|ccc|c|c|c}
\hline
\hline
Beam    & \multicolumn{3}{|c|}{Signal $\nu_e$}   & Beam $\nu_e$  & $\nu_\mu$ NC & $\nu_\tau$ CC \\
        & 0--3 GeV & 3--10 GeV & 0--10 GeV & & & \\
\hline
Baseline\cite{Stahl:2012exa} & 49 & 636 & 685 &  78           &  44        &  146         \\
GLBOPT  & 68 & 626 & 693 &  77           &  44        &  128         \\
LEOPT   & 60 & 253 & 313 &  53           &  23        &   61         \\
HEOPT   & 60 & 867 & 927 &  82           &  70        &  266         \\
\hline
\hline
\end{tabular}
\caption{Expected contributions to the e-like event sample in 0 -- 10 GeV window in the neutrino running mode, for different SPS beam optimisations for 24 kton detector size and integrated exposure of $11.25 \times 10^{20}$ POT. Additionally the signal contribution rates are shown for low energy and high energy regions. The value of $\delta_{CP} = 0$ and NH are assumed.}
\label{tab:sps_elike_rates_cp0}
\end{table}

\begin{table} [h]
\centering
\begin{tabular}{c|ccc|c|c|c}
\hline
\hline
Beam    &  \multicolumn{3}{|c|}{Signal $\nu_e$}  & Beam $\nu_e$  & $\nu_\mu$ NC & $\nu_\tau$ CC \\
        & 0--3 GeV & 3--10 GeV & 0--10 GeV & & & \\
\hline
Baseline\cite{Stahl:2012exa}  & 122 & 1589 & 1711 &  195          &  109       &  365         \\
GLBOPT  & 229 & 1361 & 1590 &  180          &  92        &  200         \\
LEOPT   & 254 & 668  & 922  &  162          &  58        &   80         \\
HEOPT   & 251 & 1546 & 1796 &  191          &  108       &  262         \\
\hline
\hline
\end{tabular}
\caption{Same as Table~\ref{tab:sps_elike_rates_cp0} but for the different HPPS beam optimisations and integrated exposure of $22.5 \times 10^{21}$ POT.}
\label{tab:hpps_elike_rates_cp0}
\end{table}

In general, the neutrino beam with lower average energy is better for the CPV sensitivity, given its enhanced coverage around
the 2nd maximum, the first minimum, and the 1st maximum. It provides the optimal sensitivity to the shape variation induced by $\delta_{CP}$. This is the main reason why the beam optimisations presented in this paper outperform the baseline LBNO beam: there is a significant increase in the number of events in those low energy regions. 

We also note that the GLB and HE optimisations result in the identical CPV sensitivity for SPS. Given comparable signal event rates at low energies, the fact that the larger rate obtained with HE option at higher energies (by almost 40\%) does not lead to substantial improvement in the performance implies that sampling oscillation probability at high energies (above 3-4 GeV) is not very efficient. The potential gain is also offset with the increasing electron background from leptonic $\tau$ decays. In the case of SPS LE, the much lower signal event rate gives a sensitivity to CPV which is not significantly worse than ones obtained with the HE and GLB optimisations. This shows that sampling the oscillation probability near the 2nd maximum leads as expected to a better sensitivity to $\delta_{CP}$. On the other hand, the low statistics leads to a worse statistical precision, which leads to a reduced CPV sensitivity.  This is not the case for HPPS LE optimisation, where the primary proton beam of 50 GeV for HPPS instead of 400 GeV for SPS, allows for significant improvement in the neutrino flux at low energies and a better coverage of the 2nd oscillation maximum. 

In summary, we conclude that the ability to access the 2nd oscillation maximum offers LBNO an opportunity to enhance the CPV sensitivity. In Section~\ref{sec:results_2ndmax}, we will give a more quantitative evaluation of the impact this region of L/E has on the measurement of the $\delta_{CP}$ parameter.

%%%%%%%%%%%%%%%%%%%%%%%%%%%%%%%%%%%%%%%%%%%%%%%%%%%%%%%%%%%%%%%%%%%%%
%             Results
%%%%%%%%%%%%%%%%%%%%%%%%%%%%%%%%%%%%%%%%%%%%%%%%%%%%%%%%%%%%%%%%%%%%%

\section{Results}
\label{sec:results}
\subsection{Overall event rates}

The expected neutrino rates for SPS GLB and HPPS LE optimisations are given in Table~\ref{tab:evrates}. The rates are computed for 24 kton detector mass and a total exposure of $15\times 10^{20}$ ($30\times 10^{21}$) POT for SPS (HPPS) beam. The fraction of the beam devoted to running with the neutrino (anti-neutrino) beam is 75\% (25\%). 

%A neutrino beam with less tail
%at high energy is better to suppress the $\tau$ production and, consequently, reduce the $\tau\rightarrow e$ events. Given comparable overall signal event rates and similar rates for the backgrounds, the optimised SPS beam gives a better performance in terms of sensitivity to CPV (see Fig.~\ref{fig:cern_beams_optim_cpv}) 
%than the baseline version which focusses more neutrinos
%at the lower energies and therefore generates more events around the second oscillation maximum. 
%In the case of HPPS, the optimised beam has a substantially lower rate of signal $\nu_e$ events. However, as can be inferred from Fig.~\ref{fig:cern_beams_optim}, these events populate mostly the energy region around and just above the 2nd maximum. This leads to the appreciable improvement in CPV sensitivity as seen in the right panel of Fig.~\ref{fig:cern_beams_optim_cpv}.

\begin{table}[h]
\begin{tiny}
\centering  
\begin{tabular}{l|c|c|c|c|ccc}
\hline\hline
                                        & $\nu_\mu$ unosc.& $\nu_\mu$ osc. &$\nu_e$ beam& $\nu_\mu\rightarrow\nu_\tau$  & \multicolumn{3}{|c}{$\nu_\mu\rightarrow\nu_e$ CC} \\
            & CC              & CC            & CC           &  CC                          & $\delta_{CP} = -\pi/2$ & 0     & $\pi/2$   \\
\hline
SPS beam, 24kton, NH                    &                &               &            &             &           &           &           \\
$11.25\times10^{20}$ POT for $\nu$       & 12492 (12397)  & 3392 (3073)   &  77 (78)   & 733 (839)   & 883 (843) & 693 (685) & 576 (558) \\
$3.75\times10^{20}$ POT for $\bar{\nu}$  & 1907 (1595)    & 504 (372)     &  10 (6)    & 112 (129)   & 32 (20)   & 40 (29)   & 41 (32)   \\
\hline
SPS beam, 24kton, IH                    &                &               &            &             &           &           &           \\
$11.25\times10^{20}$ POT for $\nu$       & 12492 (12397)  & 3337 (2929)   &  77 (78)   & 860 (918)   & 229 (233) & 138 (144) & 97 (103)  \\
$3.75\times10^{20}$ POT for $\bar{\nu}$  & 1907 (1595)    & 502 (364)     &  10 (6)    & 110 (127)   & 80 (71)   & 102 (88)  & 114 (103) \\
\hline
HPPS beam, 24kton, NH                   &                &               &            &             &            &           &           \\
$22.5\times10^{21}$ POT for $\nu$        & 19440 (30953)  & 7791 (7672)   &  162 (195) & 458 (2094)  & 1355 (2105)& 922 (1711)& 815 (1393)\\
$7.5\times10^{21}$ POT for $\bar{\nu}$   & 2073 (3982)    & 789 (930)     &  15 (16)   & 47 (322)    & 32 (49)    & 48 (71)   & 48 (80)   \\
\hline
HPPS beam, 24kton, IH                   &                &               &            &             &            &           &           \\
$22.5\times10^{21}$ POT for $\nu$        & 19440          & 7901 (7313)   &  162 (195) & 499 (2292)  & 445 (581)  & 216 (359) & 166 (257) \\
$7.5\times10^{21}$ POT for $\bar{\nu}$   & 2073 (3982)    & 801 (910)     &  15 (16)   & 46 (318)    & 84 (177)   & 122 (221) & 127 (256) \\
\hline\hline
\end{tabular}
\end{tiny}
\caption{Expected event rates in the energy range from 0 to 10 GeV for SPS and HPPS proton beam options and NH and IH. The numbers in the brackets are the event rates obtained with the baseline version of the neutrino beam. A total exposure of $15\times10^{20}$ ($30\times10^{21}$) POT is assumed for SPS (HPPS) with 75\% of running time dedicated to beam operation in the neutrino mode.}
\label{tab:evrates}
\end{table}

\begin{figure}[ht]
\begin{center}
\includegraphics[width=\the\figwidth]{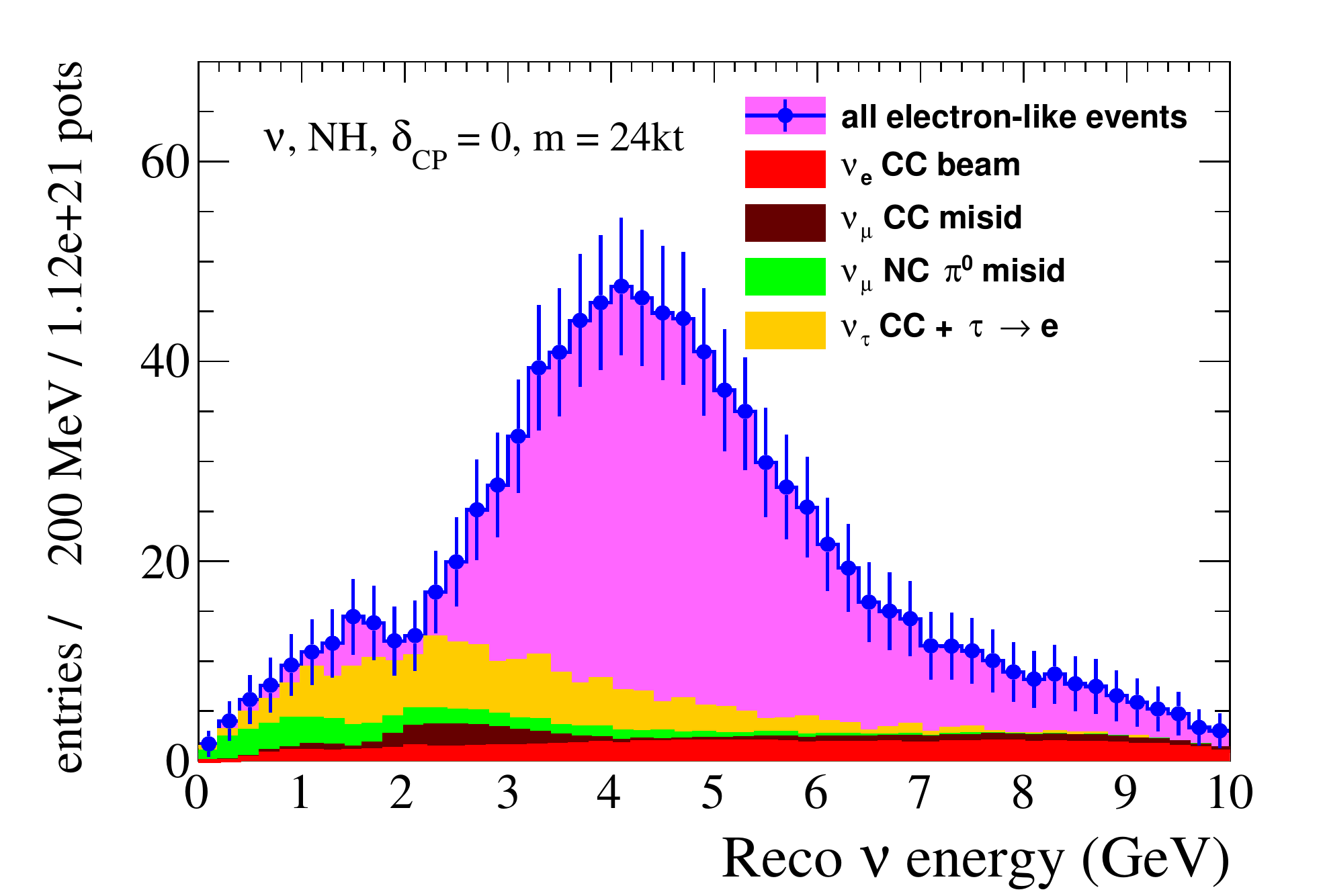}
\includegraphics[width=\the\figwidth]{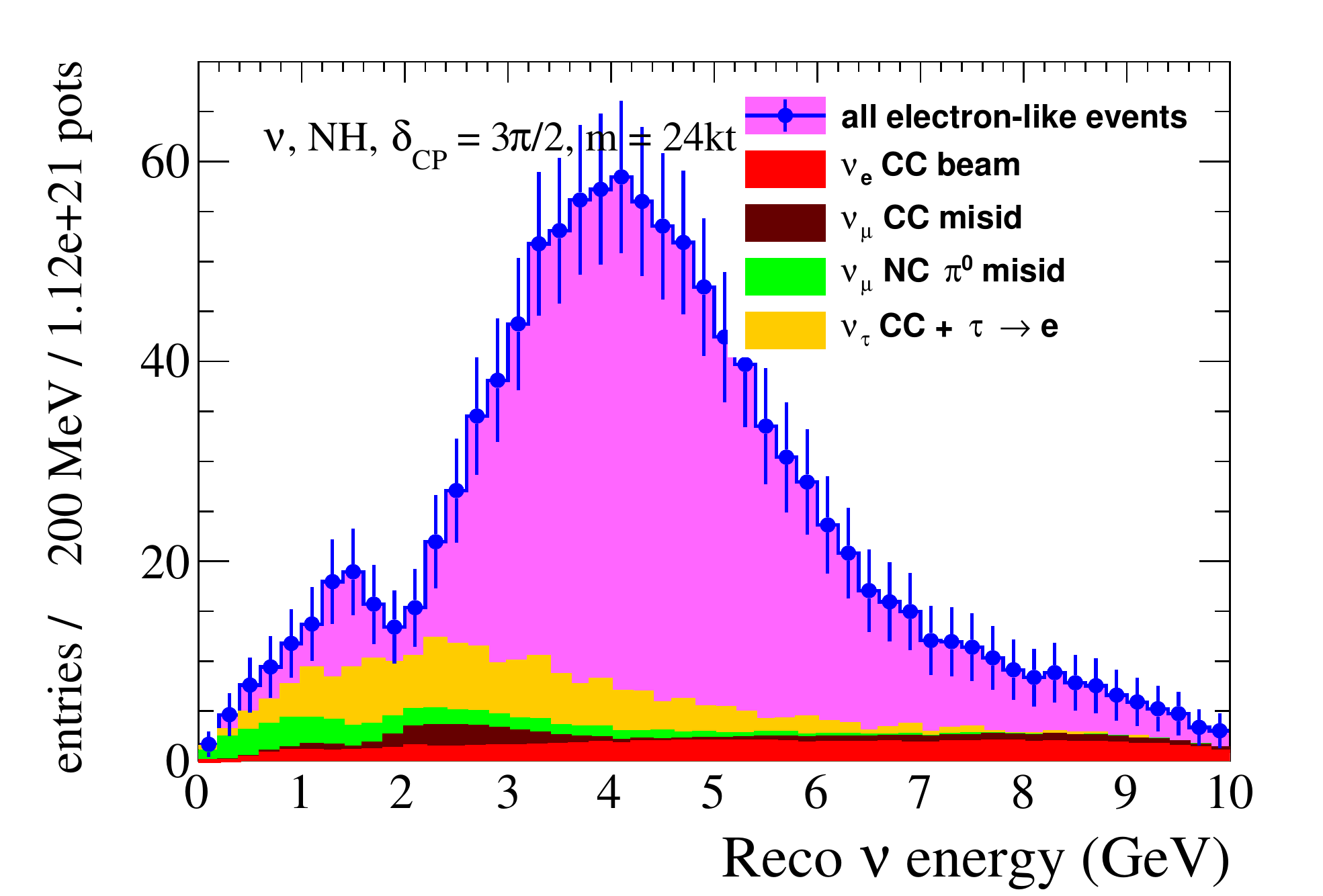}
\caption{Expected distribution of e-like events for an SPS-based neutrino beam, assuming $\delta_{CP}=0$ (left) and $\delta_{CP}=3\pi/2$ (right) and NH. The error bars represent statistical uncertainties only.}
\label{fig:nue_eventdist_sps}
\end{center}
\end{figure}

\begin{figure}[ht]
\begin{center}
\includegraphics[width=\the\figwidth]{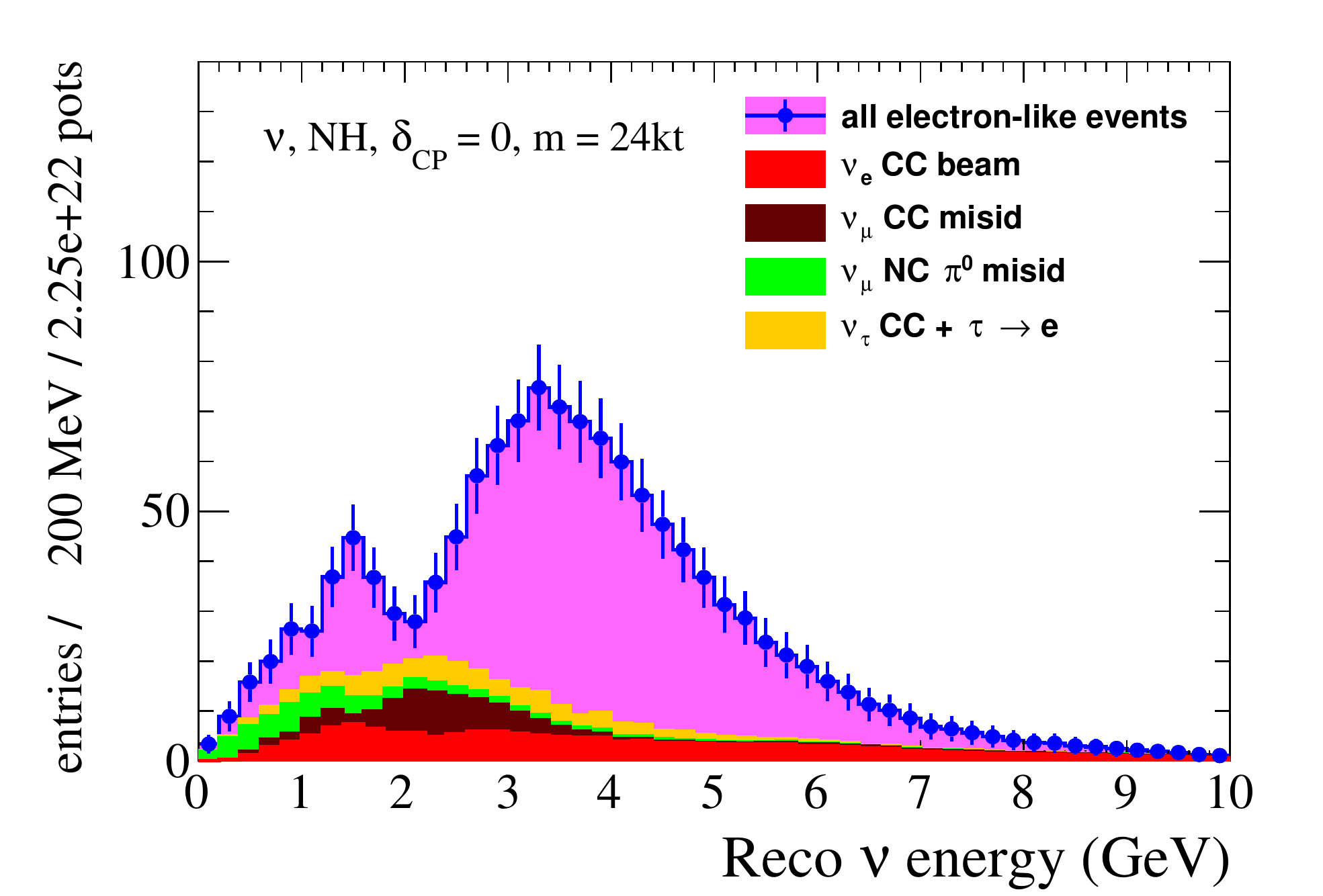}
\includegraphics[width=\the\figwidth]{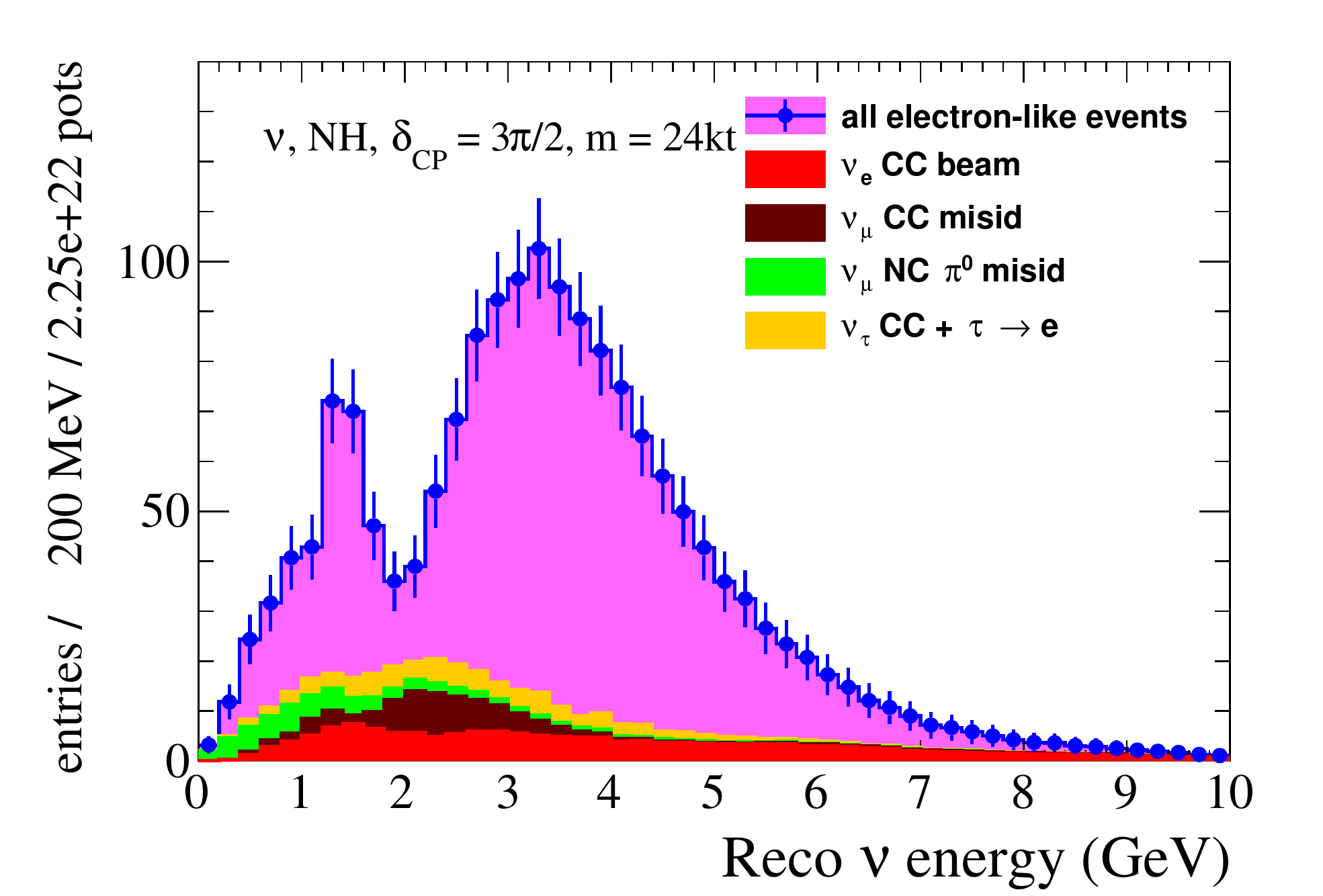}
\caption{Expected distributions of e-like events for HPPS-based neutrino beam operating assuming $\delta_{CP}=0$ (left) and $\delta_{CP}=3\pi/2$ (right) and NH. The error bars represent statistical uncertainties only.}
\label{fig:nue_eventdist_hpps}
\end{center}
\end{figure}

\begin{figure}[ht]
\begin{center}
\includegraphics[width=\the\figwidth]{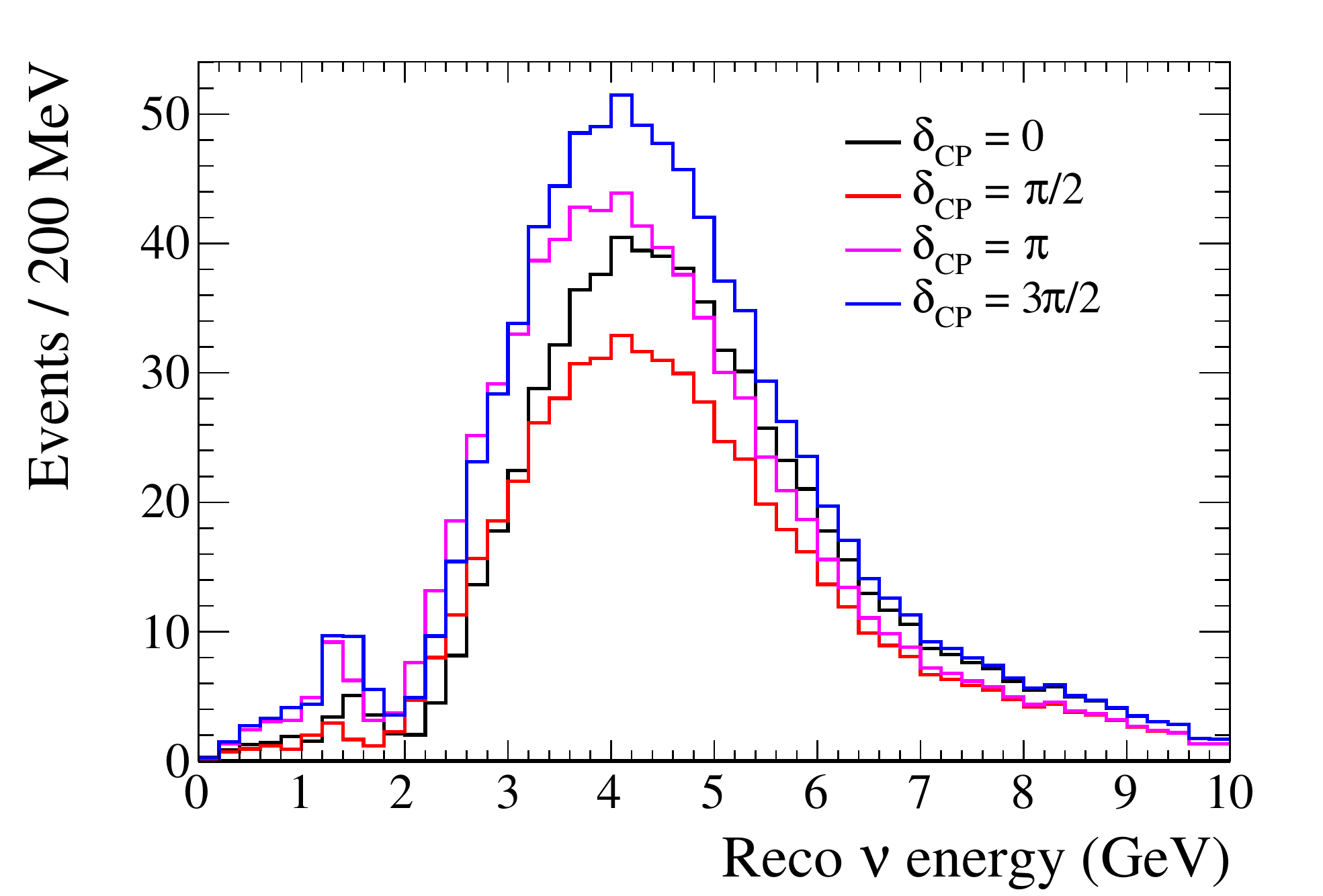}
\includegraphics[width=\the\figwidth]{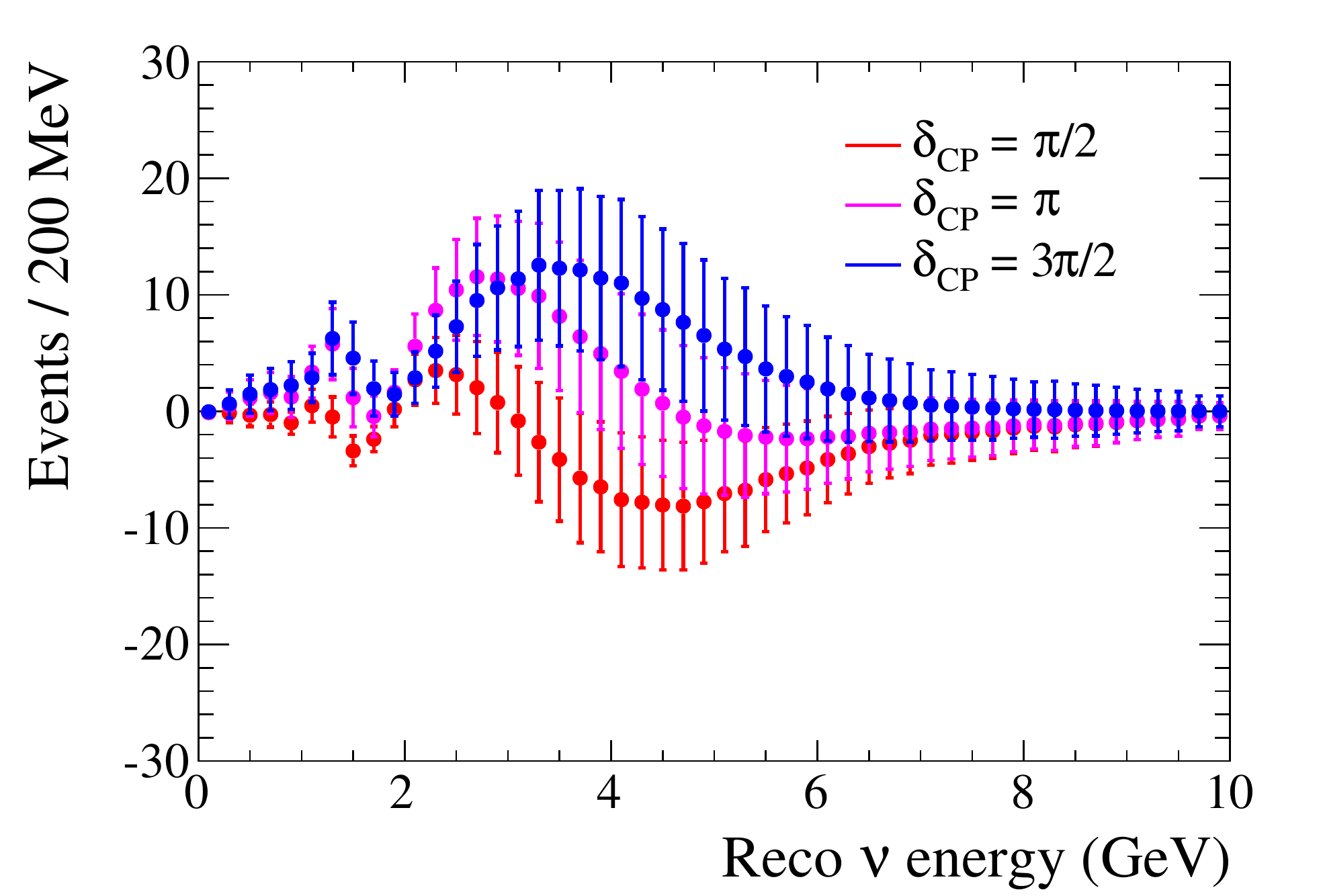}
\caption{Left: Distribution of signal e-like events in $\nu$ mode for optimised SPS beam and 24kton detector and 75\% of the total exposure of $15\times10^{20}$ POT. Right: Difference of the reconstructed energy distribution of the e-like signal events from the case of $\delta_{CP} = 0$. The error bars represent the statistical uncertainty for each energy bin. NH is assumed.}
\label{fig:nuesig_nh_eventdist_sps}
\end{center}
\end{figure}

\begin{figure}[ht]
\begin{center}
\includegraphics[width=\the\figwidth]{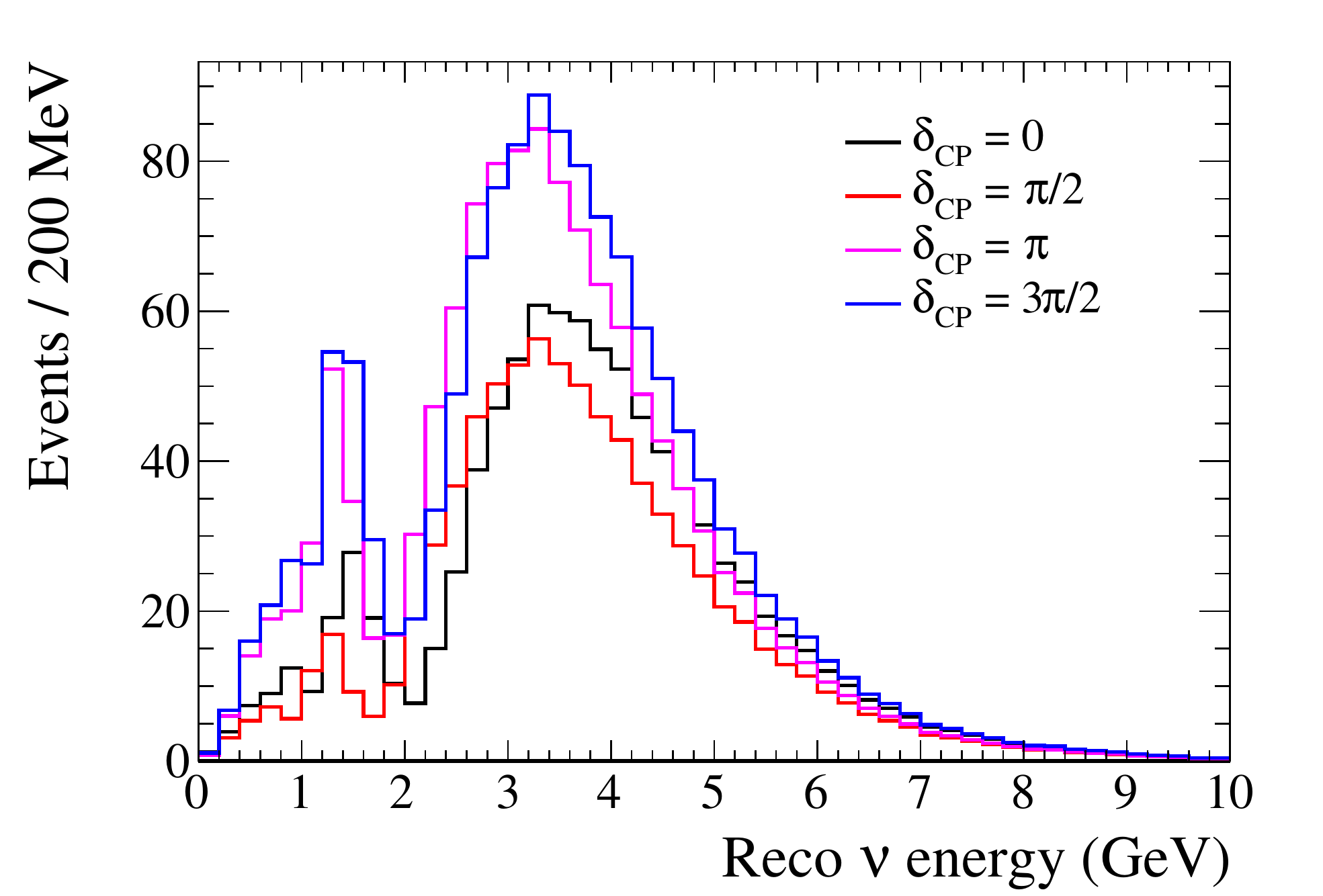}
\includegraphics[width=\the\figwidth]{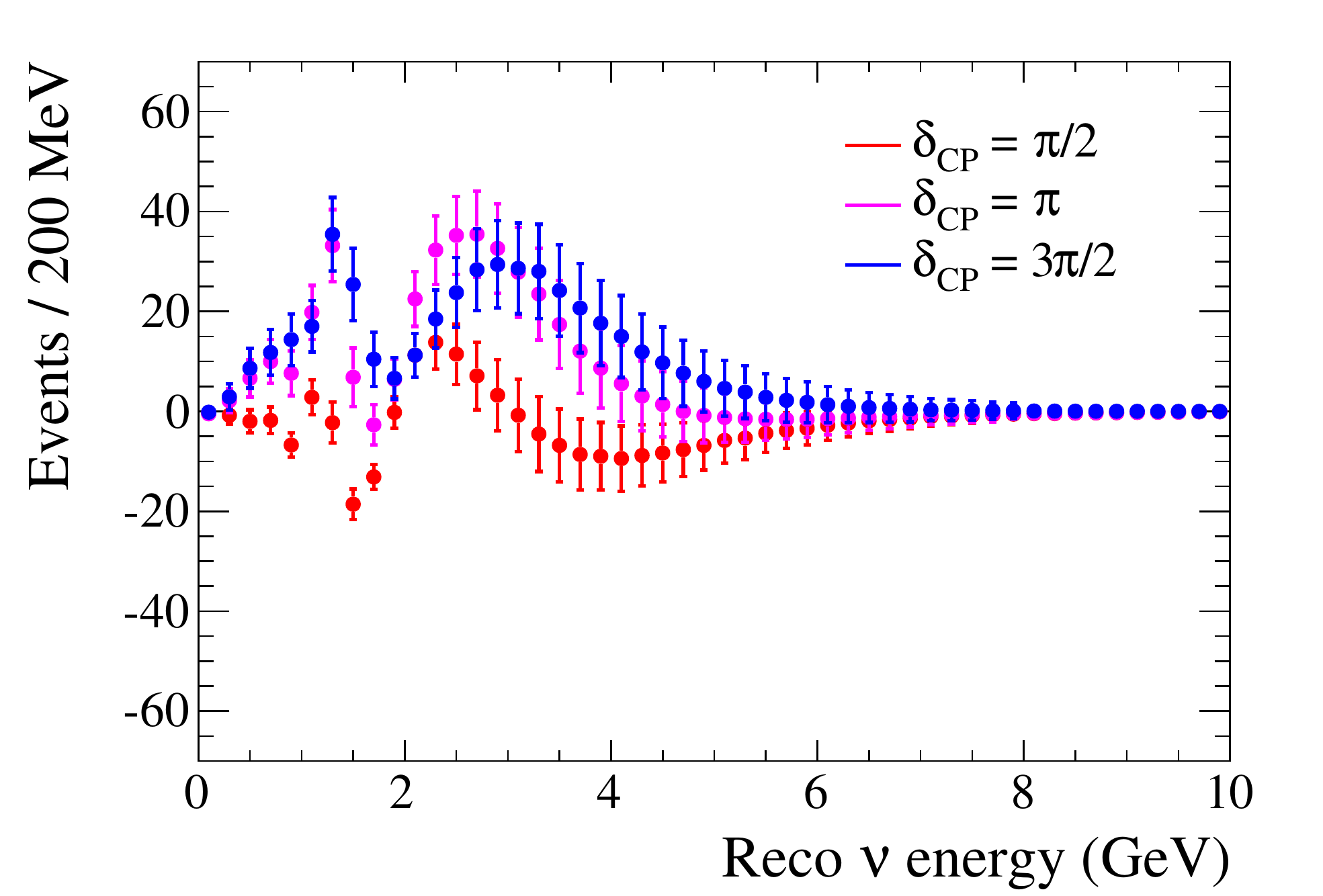}
\caption{Left: Distribution of signal e-like events in $\nu$ mode for optimised HPPS beam and 24kton detector and 75\% of the total POT exposure of  $30\times10^{21}$. Right: Difference of the reconstructed energy distribution of the e-like signal events from the case of $\delta_{CP} = 0$. The error bars represent the statistical uncertainty for each energy bin. NH is assumed.}
\label{fig:nuesig_nh_eventdist_hpps}
\end{center}
\end{figure}

The expected distributions for e-like events as a function of reconstructed neutrino energy are shown in Figure~\ref{fig:nue_eventdist_sps} and Figure~\ref{fig:nue_eventdist_hpps} for SPS and HPPS beams, respectively. Two cases are illustrated that of $\delta_{CP}=0$ and $\delta_{CP}=3\pi/2$ for each proton beam option. A further illustration of the spectral information carried by $\nu_e$ signal events is given in Figure~\ref{fig:nuesig_nh_eventdist_sps} and Figure~\ref{fig:nuesig_nh_eventdist_hpps} for SPS and HPPS beam, respectively. The energy distributions of the signal-only events under several $\delta_{CP}$ hypotheses are shown on the left, while on the right the difference with respect to the case of $\delta_{CP} = 0$ is displayed. The 2nd oscillation maximum is clearly visible in the energy window between 1 and 2~GeV for the HPPS beam. Even in the case of the SPS beam, the spectral information contained in the events around the 2nd peak is significant (see Section~\ref{sec:results_2ndmax}).

\subsection{Sensitivity to MH}
Quick and conclusive determination of MH during the first few years of operation is one of the main objectives of the LBNO physics program in the first phase of the experiment. It was shown in Ref.~\cite{Agarwalla::2013kaa} that LBNO has close to 100\% statistical power to select a correct MH hypothesis independent of the value of $\delta_{CP}$ at 3$\sigma$ level after accumulating $2\times 10^{20}$ POT or about two years of SPS beam operation. With this exposure one also achieves more than 60\% probability of determining MH at 5$\sigma$ level. The discovery (at 5$\sigma$ level) can essentially be guaranteed with an exposure of $4\times 10^{20}$ POT or about four years of running.

\begin{figure}[h]
\begin{center}
\includegraphics[width=\the\figwidth]{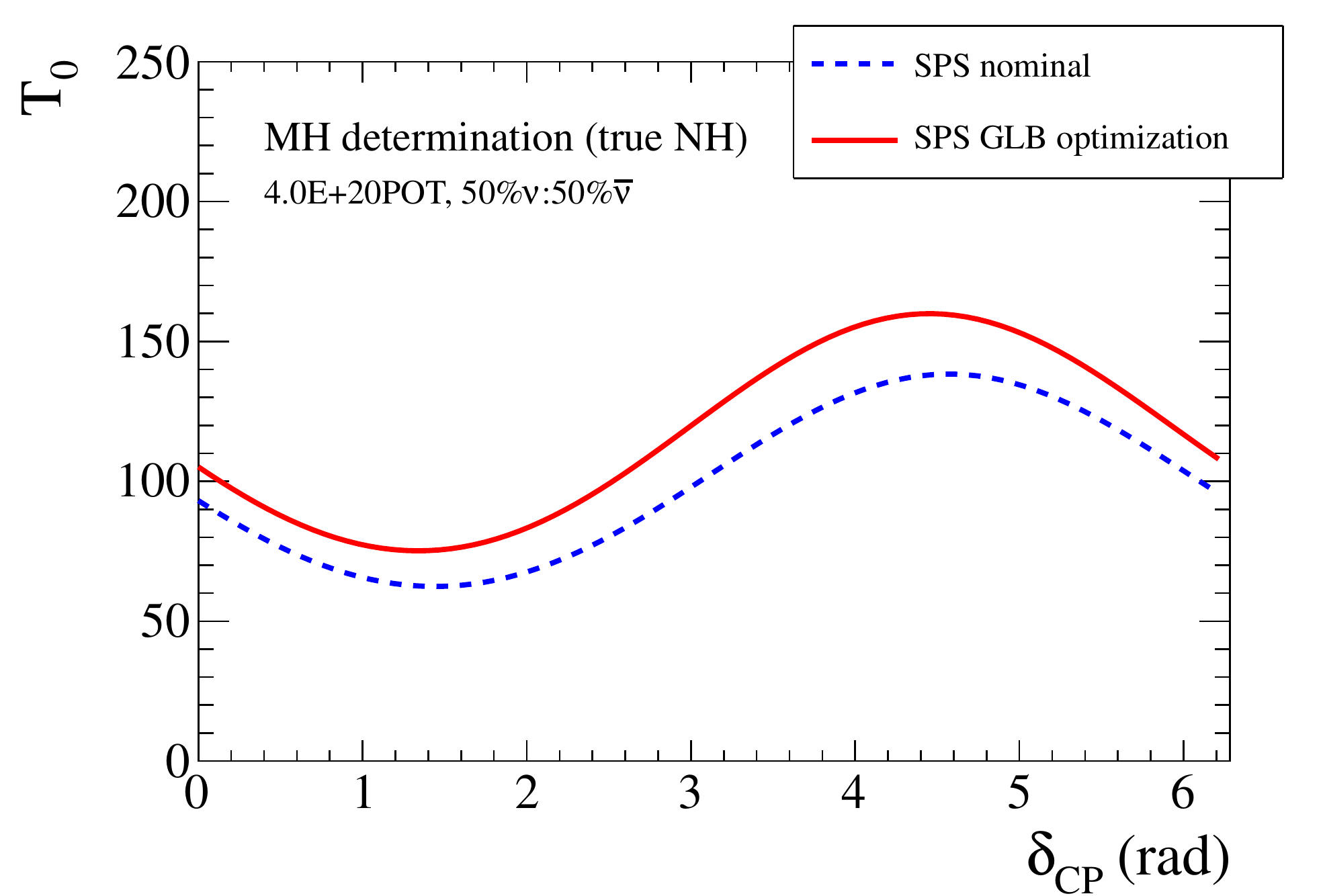}
\includegraphics[width=\the\figwidth]{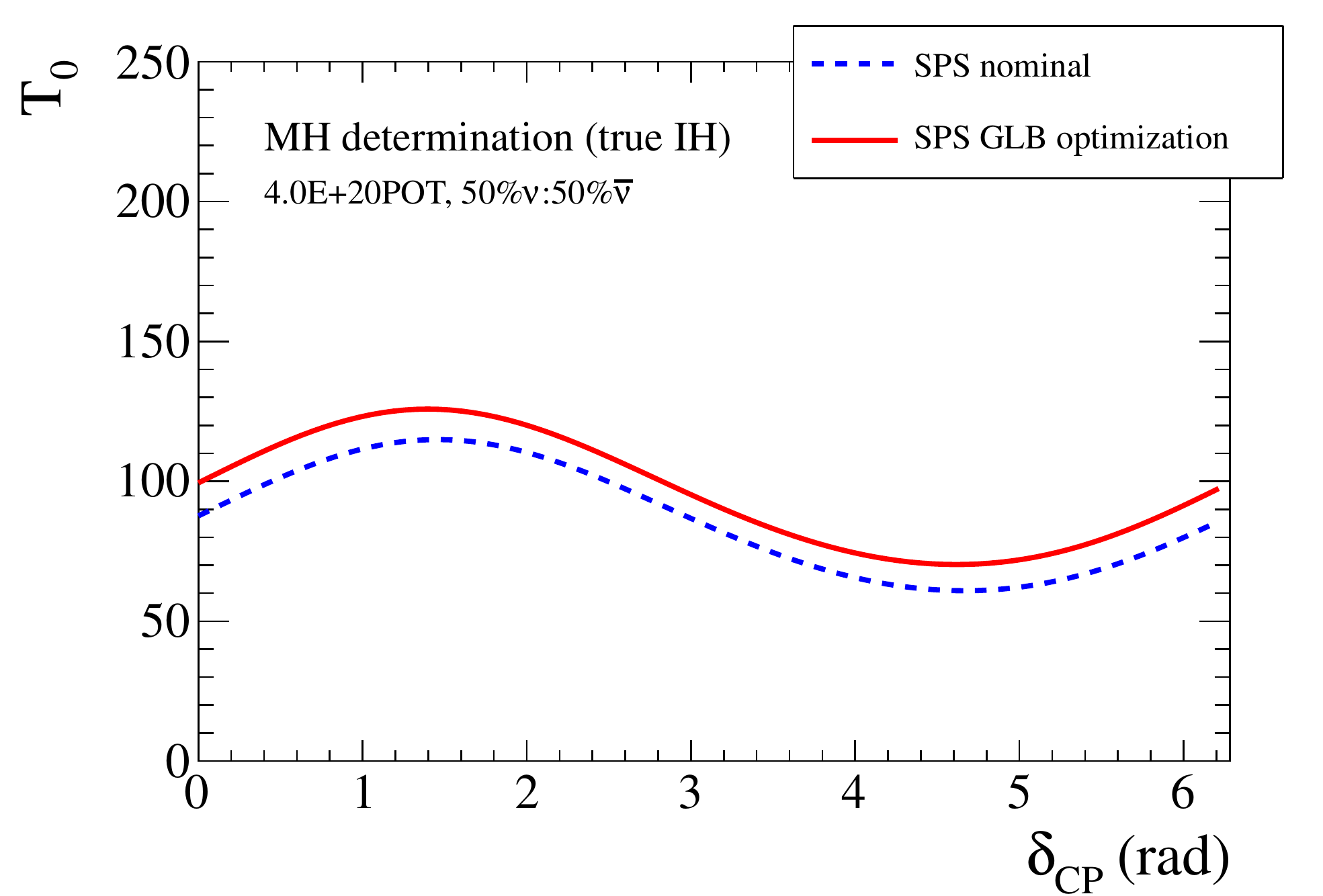}
\caption{Mean value of the mass hierarchy test statistic for nominal and optimised SPS neutrino beams as a function of true $\delta_{CP}$ value for an exposure of $4\times 10^{20}$ POT (or about 4 years of running with SPS) and 24 kton detector.}
\label{fig:T0_mh_sps}
\end{center}
\end{figure}

While the focus of the beam optimisation has been on the improvement of the LBNO sensitivity to the CPV effects, it is important to check that neutrino fluxes found to be optimal for CPV measurement do not compromise the discovery potential for MH. To examine the MH sensitivity, we define a test statistic $T$:
\begin{equation}
\label{eq:mhtestT}
T=\chi^{2}_{IH} - \chi^{2}_{NH},
\end{equation}
where $\chi^2_{IH}$ ($\chi^2_{NH}$) is obtained by minimising the $\chi^2$ in Eq.~\ref{eq:totchi2} with respect to the systematic and oscillation parameters (including $\delta_{CP}$ around the negative (positive) value of $\Delta m^2_{31}$. A comparison of the mean values for the test statistic, $T_0$, obtained with the nominal and optimised SPS neutrino beams is shown in Figure~\ref{fig:T0_mh_sps} for both NH and IH. The values of $T_{0}$ are larger for the optimised beam than for the nominal one. The sensitivity to MH is, therefore, better with the optimised neutrino flux. 

\subsection{Sensitivity to CPV}

\begin{figure}[th]
\begin{center}
\includegraphics[width=\the\figwidth]{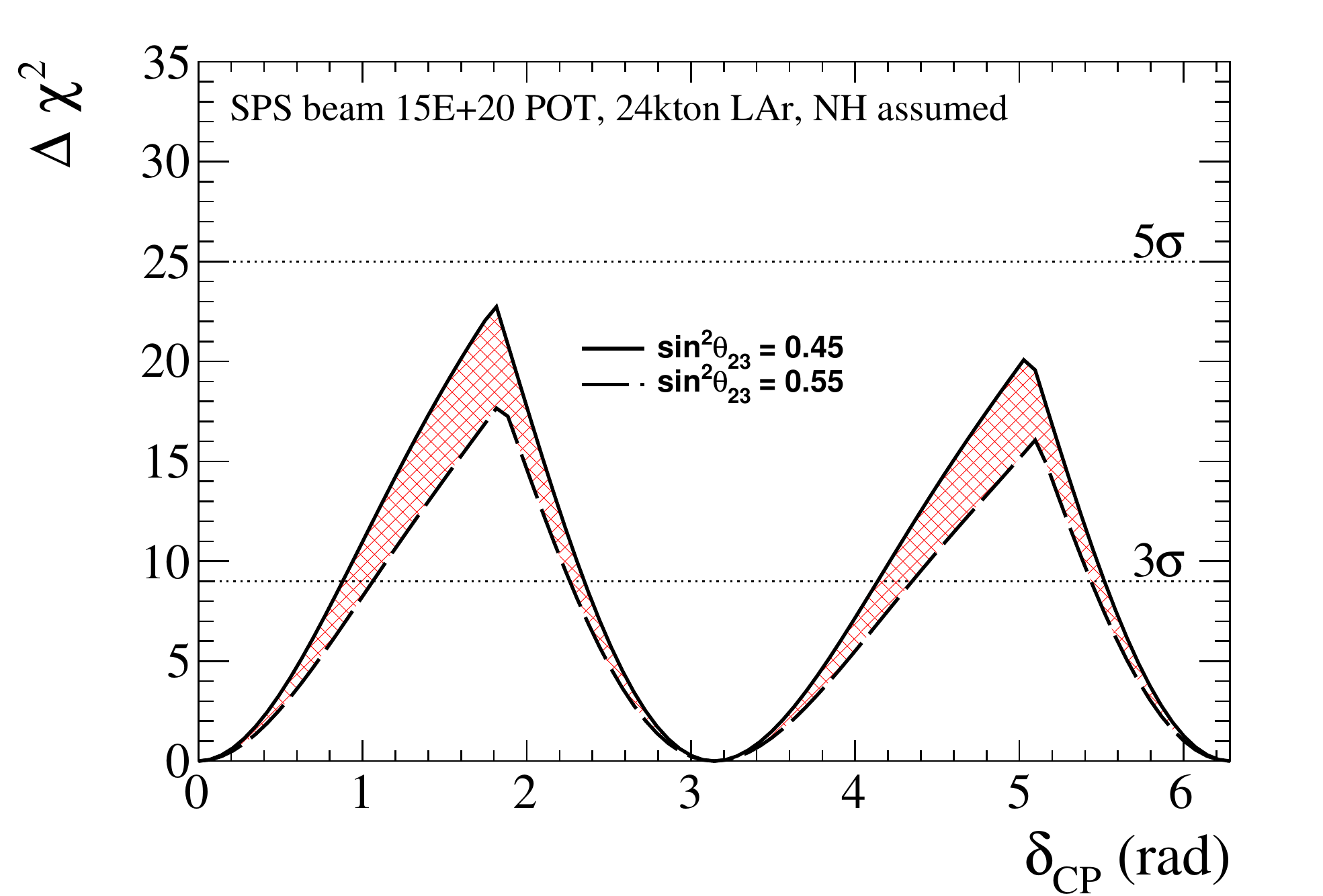}
\includegraphics[width=\the\figwidth]{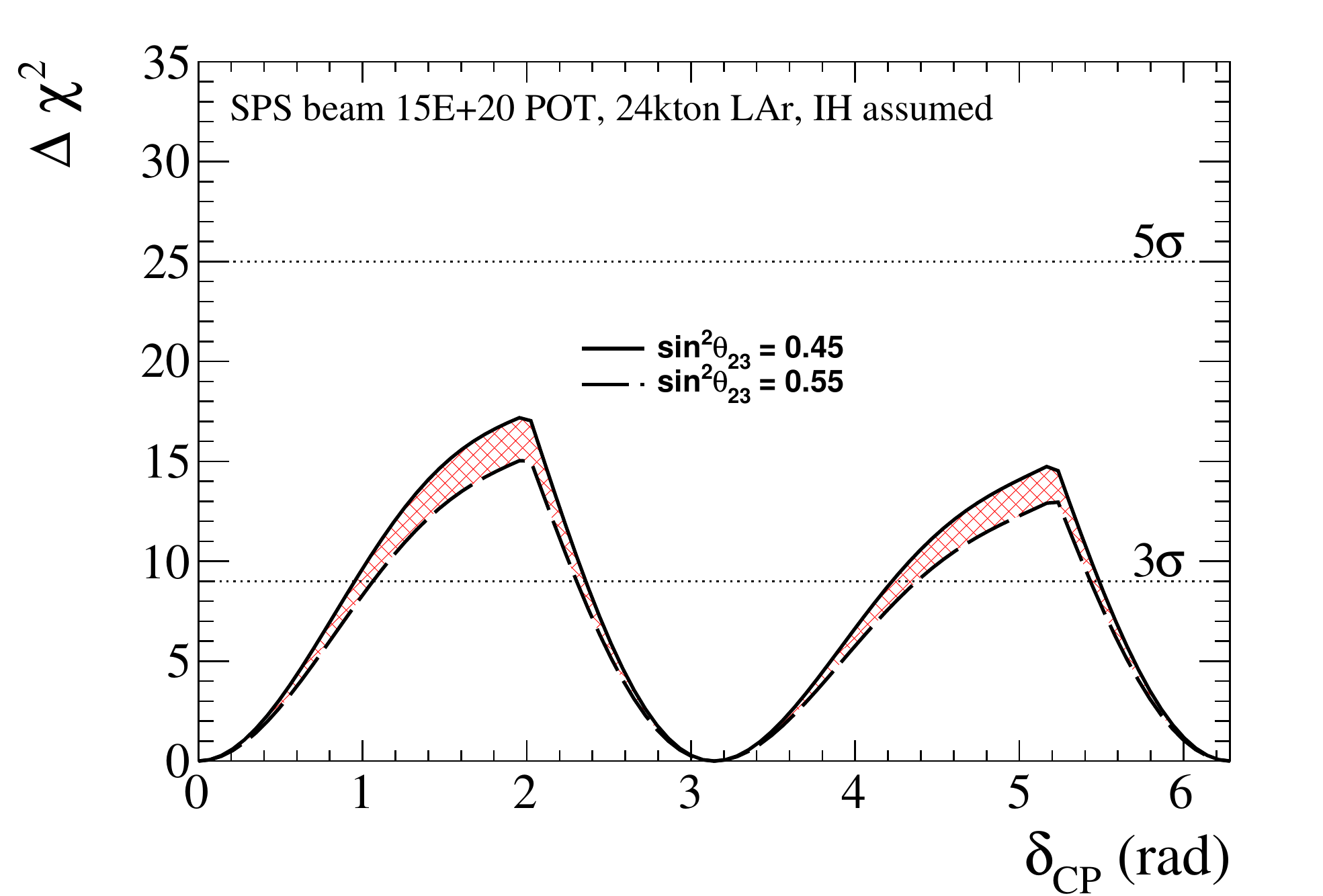}
\caption{Median sensitivity to CPV for the optimised SPS beam and LBNO20. The sensitivity is shown for a range of values of $\sin^2{\theta_{23}}$ between 0.45 and 0.55. The case of NH is shown on the left, while that of IH is shown on the right.}
\label{fig:chi2_cpv_spsopt_th23dep}
\end{center}
\end{figure}

The assumption on $\theta_{23}$ influences the sensitivity to CPV with a smaller value of $\sin^2{\theta_{23}}$ leading to a better precision in the measurement of $\delta_{CP}$ phase. In Figure~\ref{fig:chi2_cpv_spsopt_th23dep}, we show the effect the variation of $\sin^2{\theta_{23}}$ between 0.45 and 0.55 (see Table~\ref{tab:oscparam_may2014} and related discussion) on the projected senstivity to CPV. Depending on the value of $\sin^2{\theta_{23}}$, the coverage at $3\sigma$ level of $\delta_{CP} \neq 0,\pi$ parameter space varies from 45\% (43\%) to 36\% (36\%) for the case of NH (IH).

\begin{figure}[htb]
\begin{center}
\includegraphics[width=\the\figwidth]{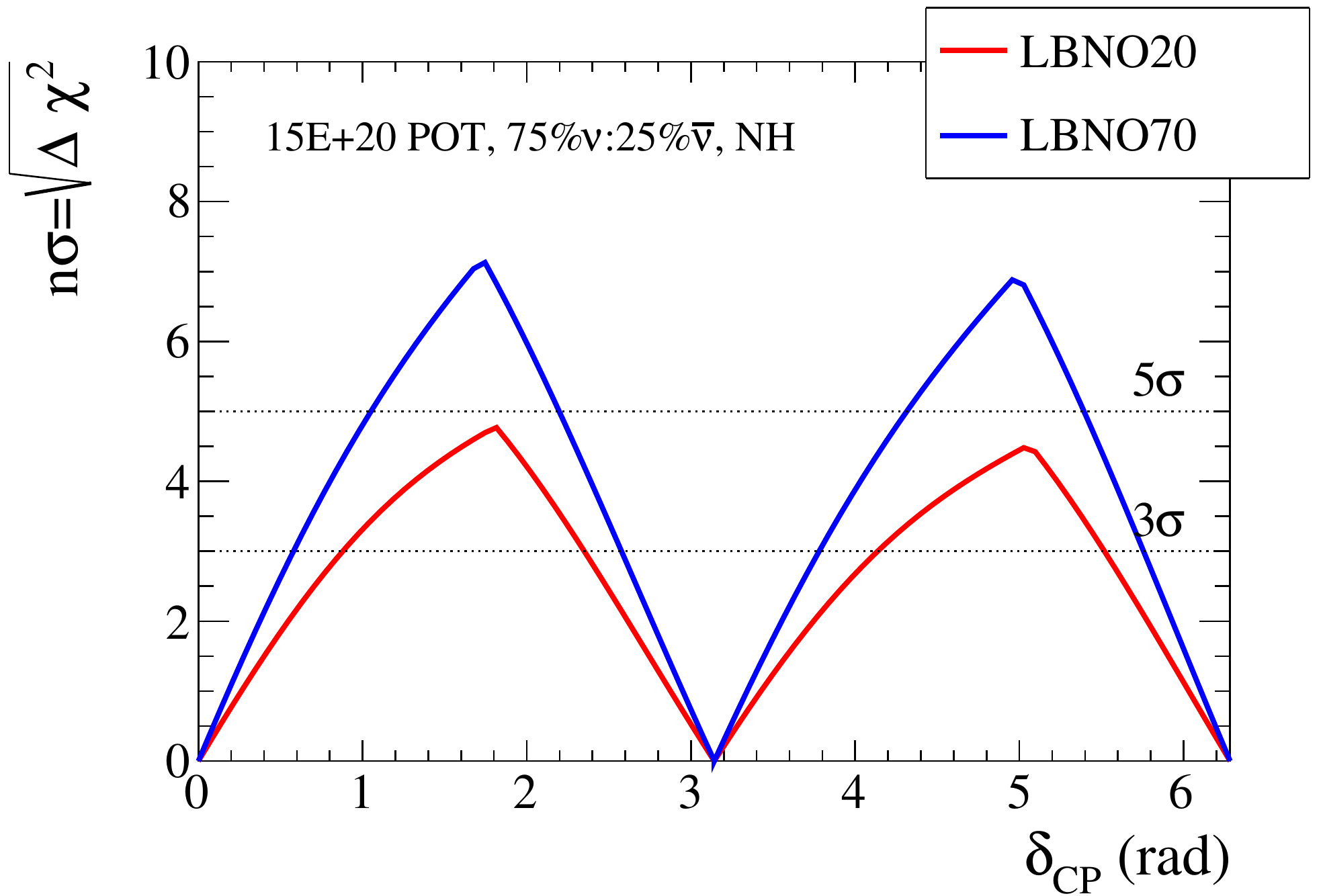}
\includegraphics[width=\the\figwidth]{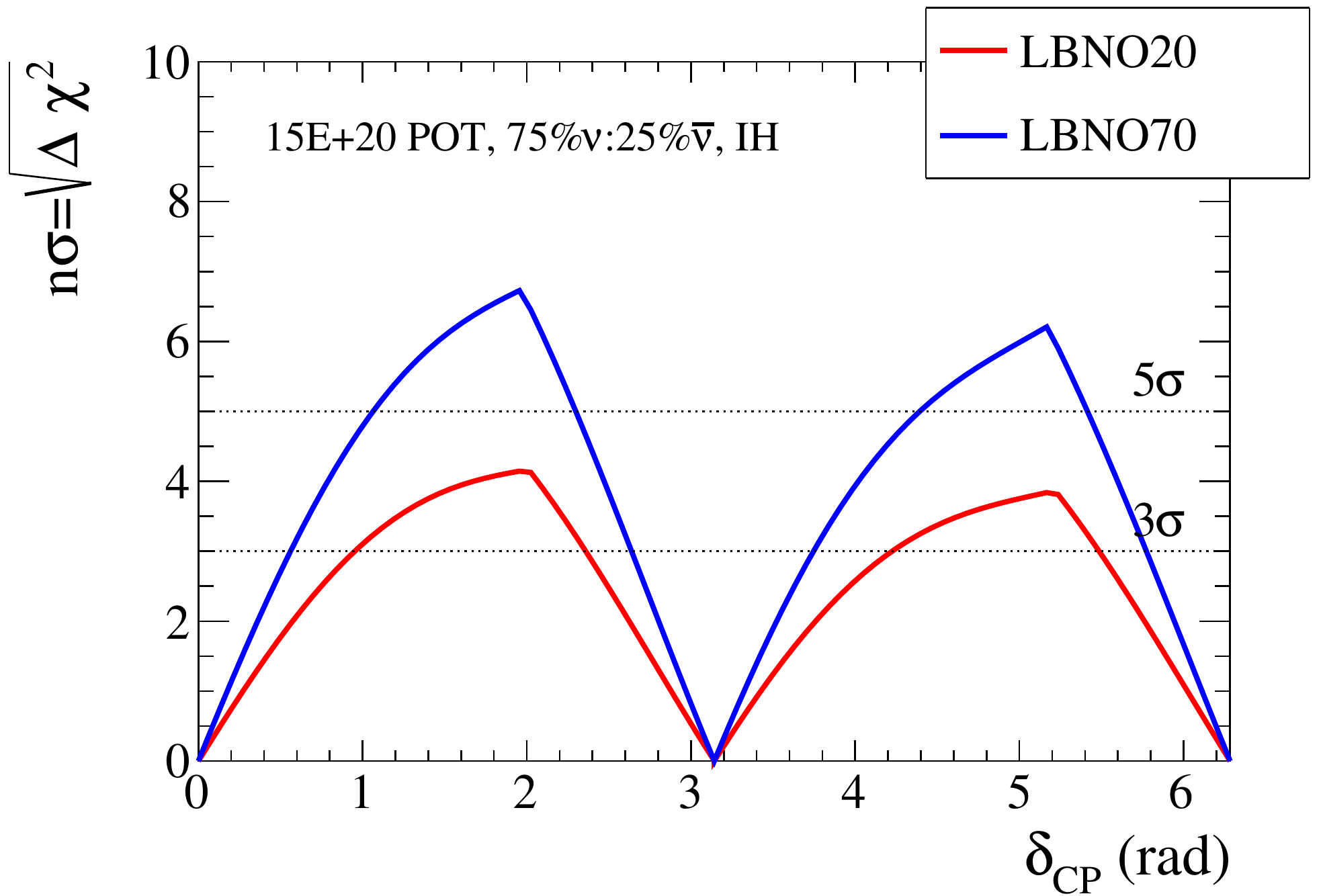}
\caption{Median sensitivity to CPV for the optimised SPS beam. The case of NH is shown on the left, while that of IH is shown on the right. The value of $\sin^2{\theta_{23}} = 0.45$ is assumed.}
\label{fig:nsig_cpv_spsopt}
\end{center}
\end{figure}

\begin{figure}[tb]
\begin{center}
\includegraphics[width=\the\figwidth]{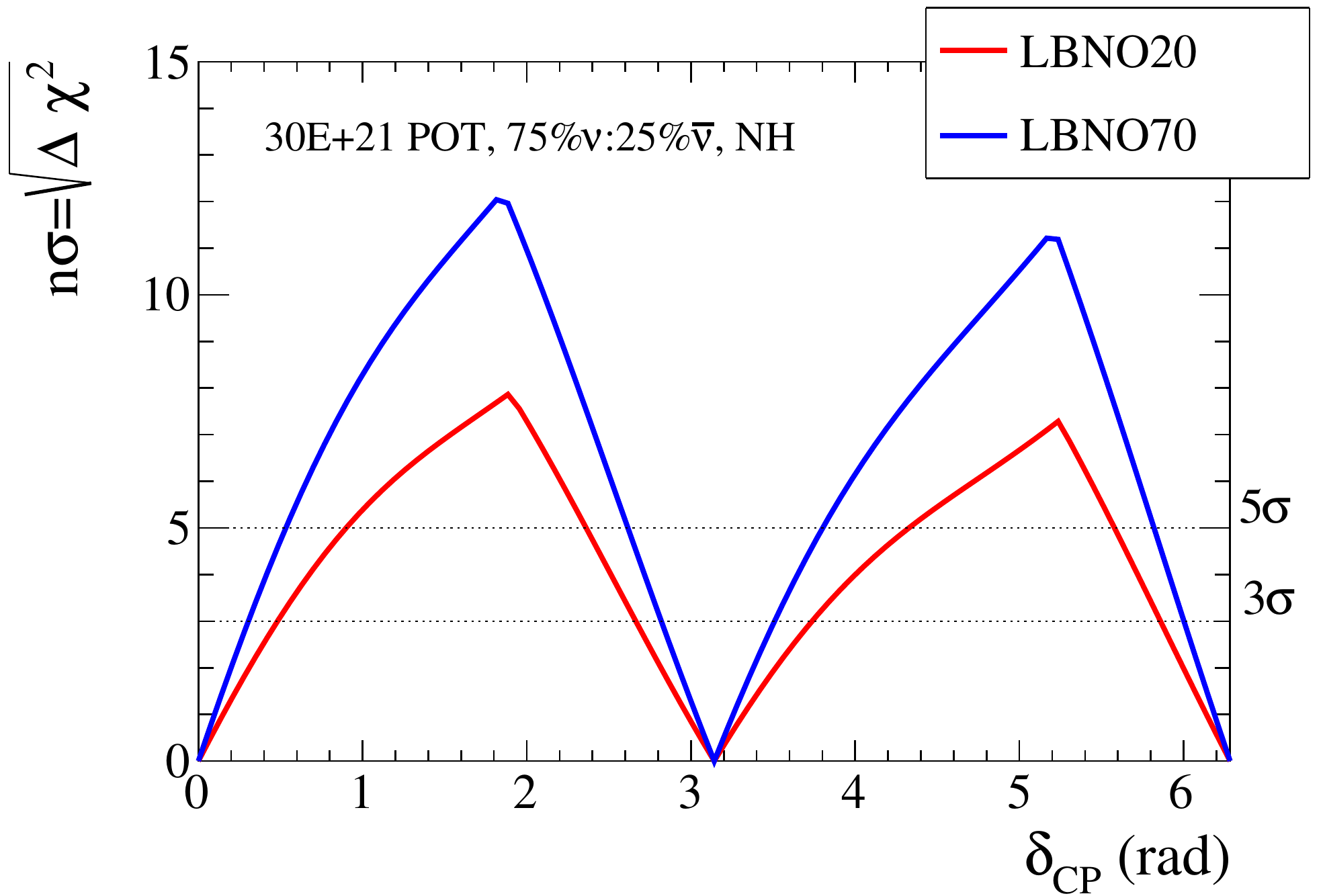}
\includegraphics[width=\the\figwidth]{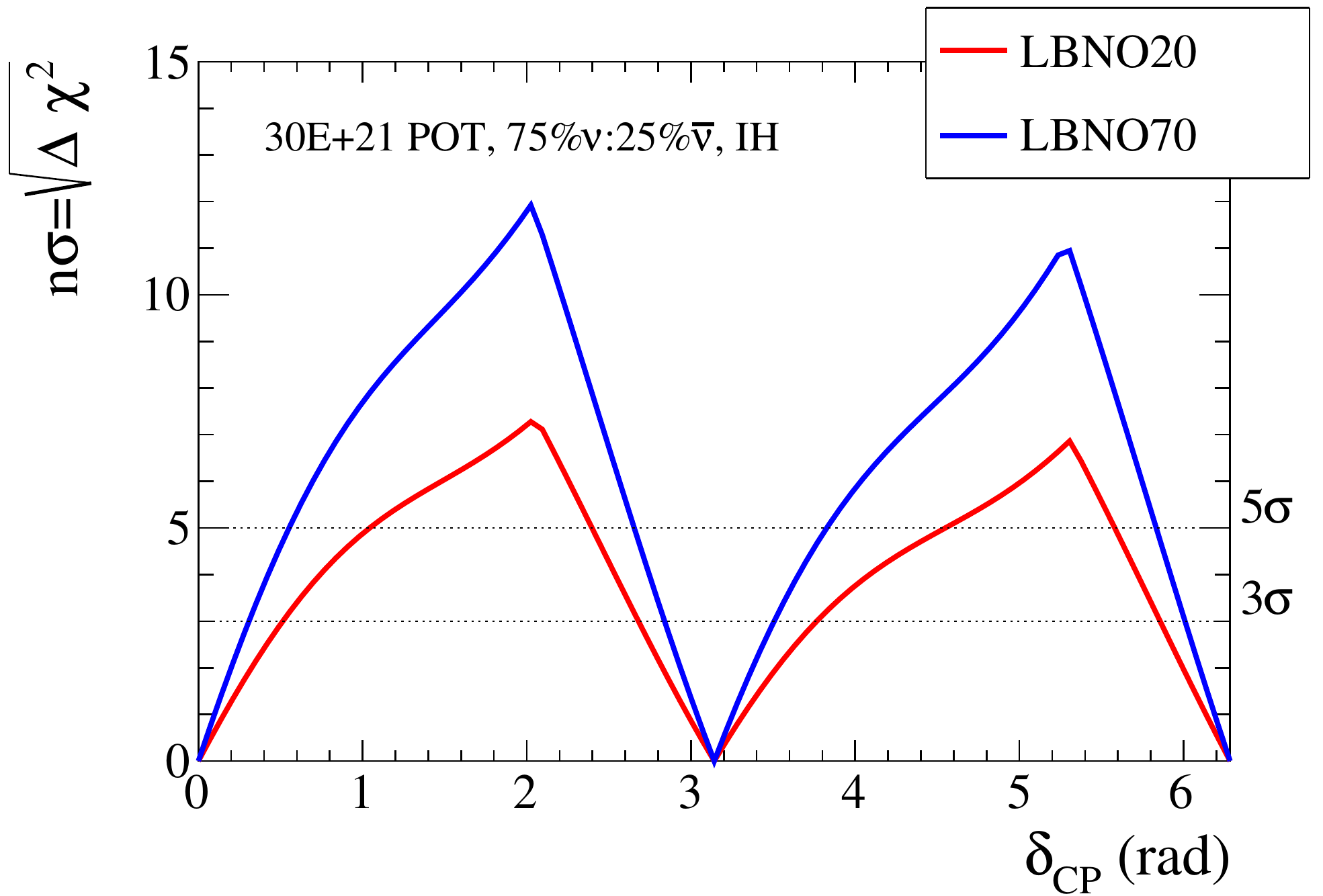}
\caption{Median sensitivity to CPV for the optimised HPPS beam. The case of NH is shown on the left, while that of IH is shown on the right. The value of $\sin^2{\theta_{23}} = 0.45$ is assumed.}
\label{fig:nsig_cpv_hppsopt}
\end{center}
\end{figure}

\begin{figure}[tb]
\begin{center}
\includegraphics[width=\the\figwidth]{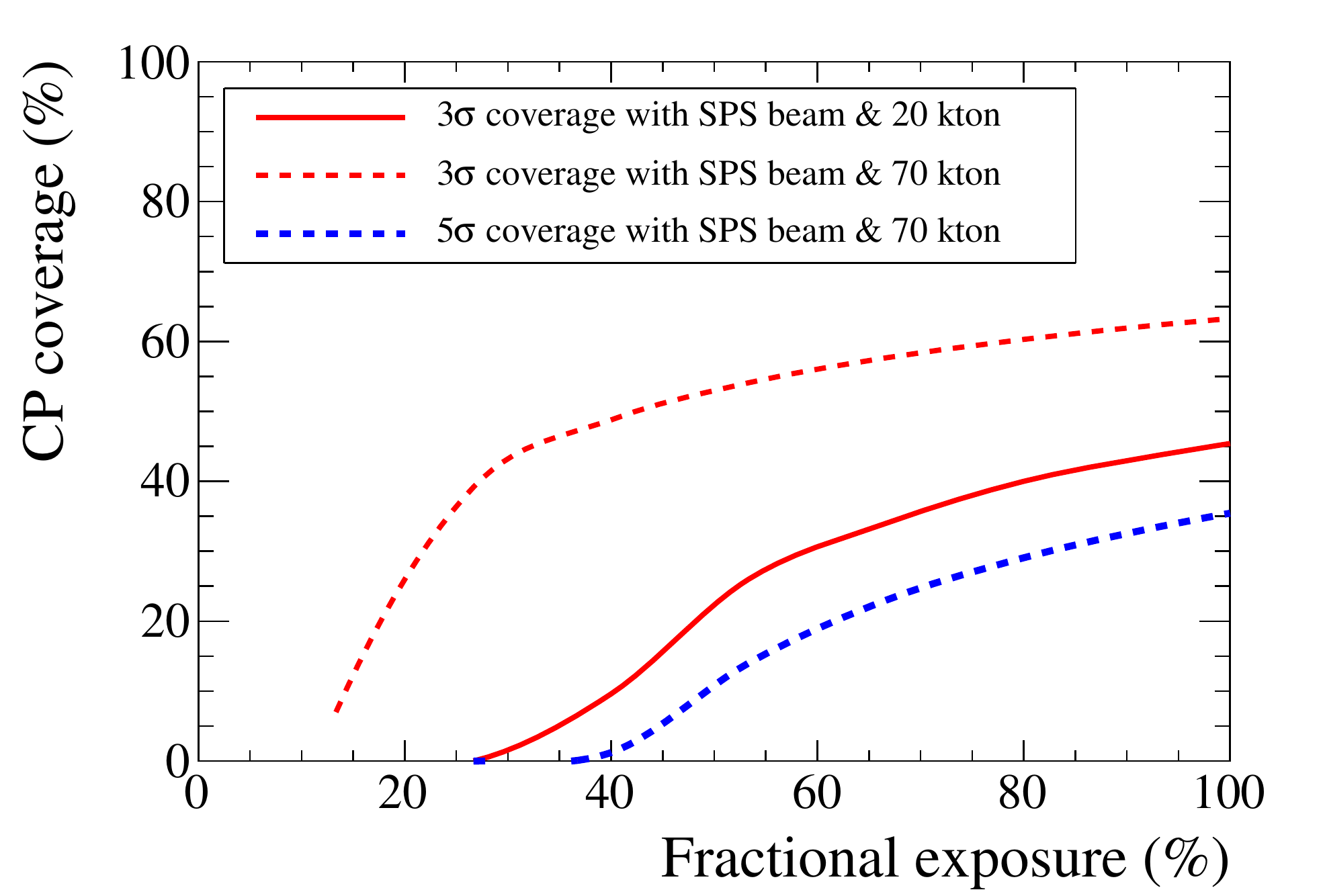}
\includegraphics[width=\the\figwidth]{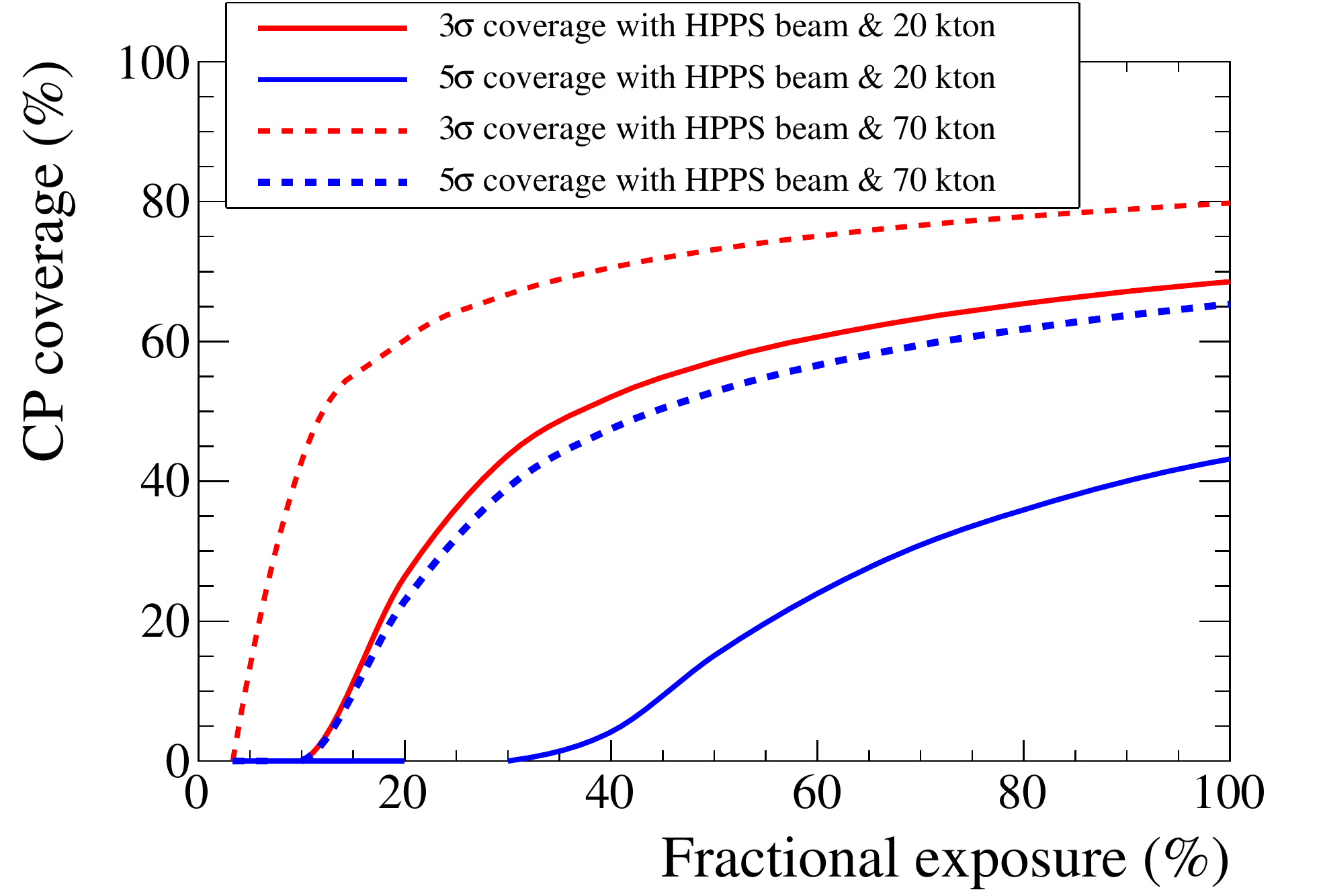}
\caption{Fractional coverage of $\delta_{CP}$ parameter space at $3\sigma$ and $5\sigma$ level with SPS (left) and HPPS (right) based neutrino beams and two detector size options. The value of $\sin^2{\theta_{23}} = 0.45$ is assumed.}
\label{fig:frac_cpvcover}
\end{center}
\end{figure}

The sensitivity of LBNO to CPV is shown in Figure~\ref{fig:nsig_cpv_spsopt} for the SPS beam option for both cases of NH and IH. Both detector mass option, 24 kton and 70kton, are displayed. Similarly, Figure~\ref{fig:nsig_cpv_hppsopt} shows the sensitivity to CPV obtained with HPPS beam, that is envisaged as one possible upgrade for the second phase of LBNO. The fractional coverage of $\delta_{CP}$ parameter space at $3\sigma$ and $5\sigma$ levels for both beam options and detector masses can be seen in Fig~\ref{fig:frac_cpvcover}. In Phase I (LBNO20: SPS based neutrino beam and 24 kton LAr detector), LBNO has a median sensitivity to CPV at or above $3\sigma$ level for about 45\% of possible values of $\delta_{CP}$. With addition of the 50 kton detector, the experiment would be able to cover 63\% and 35\% of the parameter space at $3\sigma$ and $5\sigma$ level, respectively. Ultimately, the combination of HPPS and 70 kton detector would allow to reach 80\% coverage at $3\sigma$ level, while being able to make a discovery of CPV for 65\% values of the $\delta_{CP}$ phase.

%%%%%%%%%%%%%%%%%%%%%%%%%%%%%%%%%%%%%%%%%%%%%%%%%%%%%%%%%%%%%%%%%%%%%%%%%%%%%%%%%%%%%%%%%
\subsection{Impact of the second oscillation maximum}
\label{sec:results_2ndmax}

\begin{figure}[ht]
\begin{center}
\includegraphics[width=\the\figwidth]{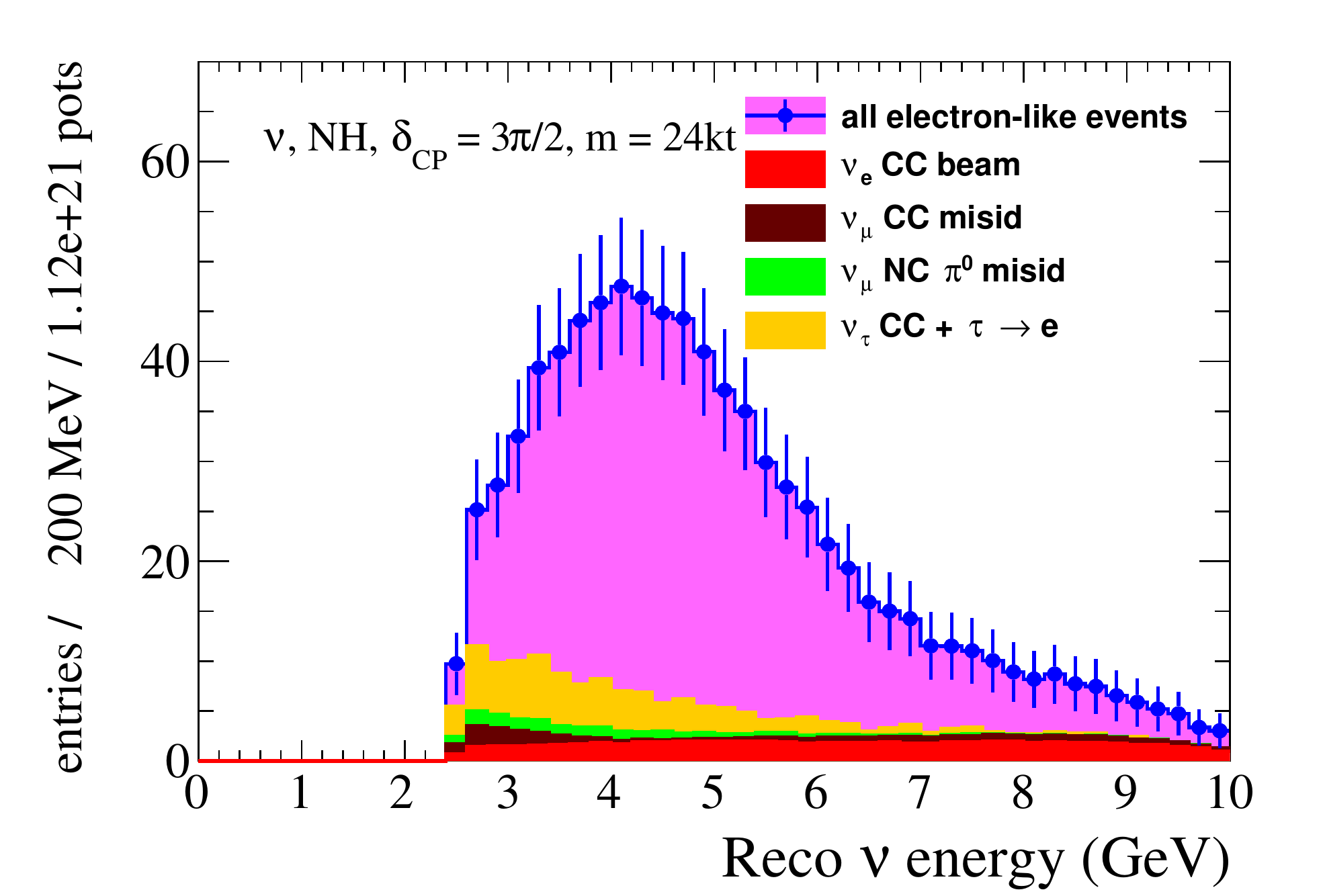}
\includegraphics[width=\the\figwidth]{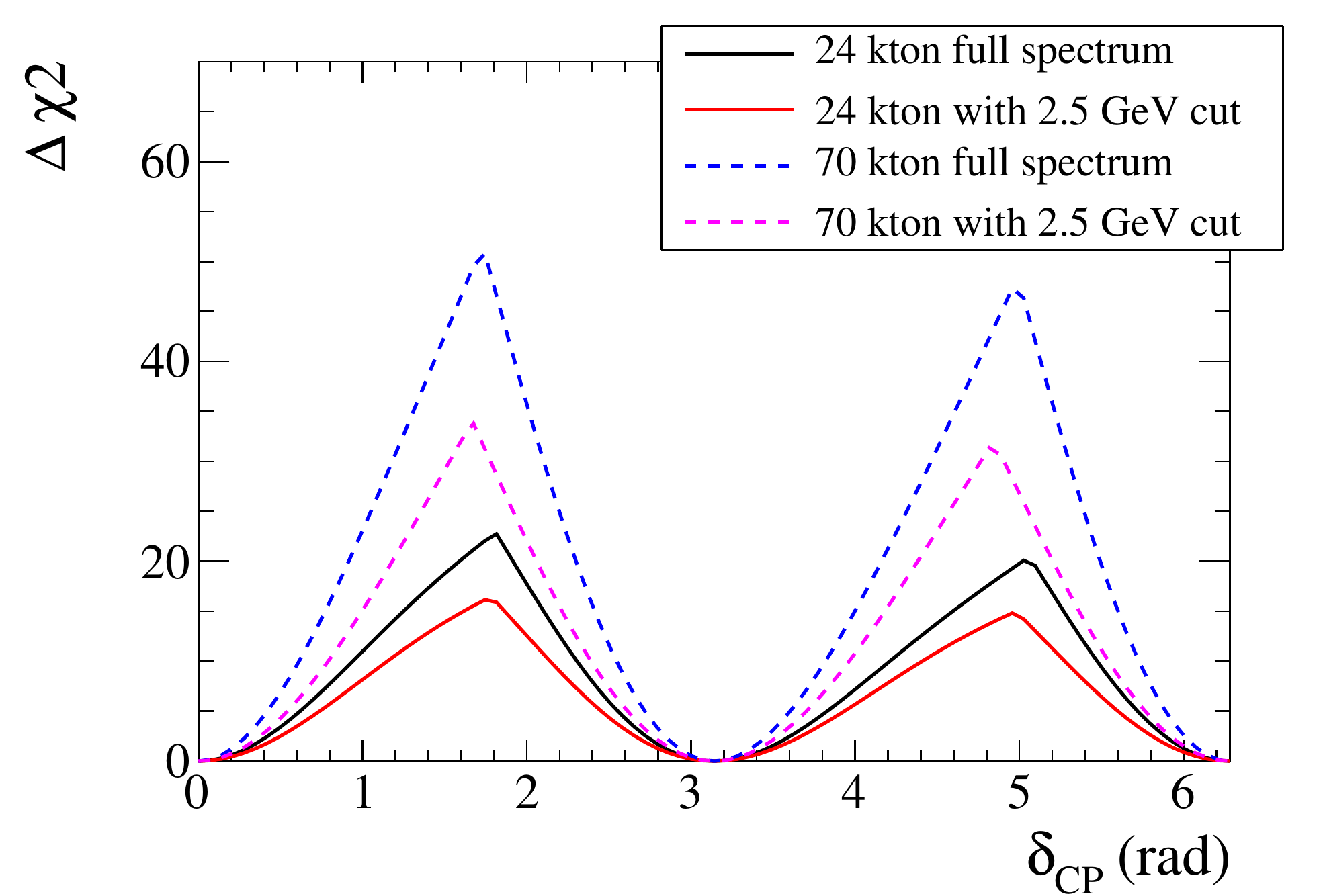}
\caption{Left: Expected distribution of e-like events for SPS-based neutrino beam after a cut on reconstructed energy at 2.5 GeV. Right: Comparison of the sensitivity to CPV with and without the energy cut.}
\label{fig:sps_2ndmax_ecut}
\end{center}
\end{figure}

\begin{figure}[ht]
\begin{center}
\includegraphics[width=\the\figwidth]{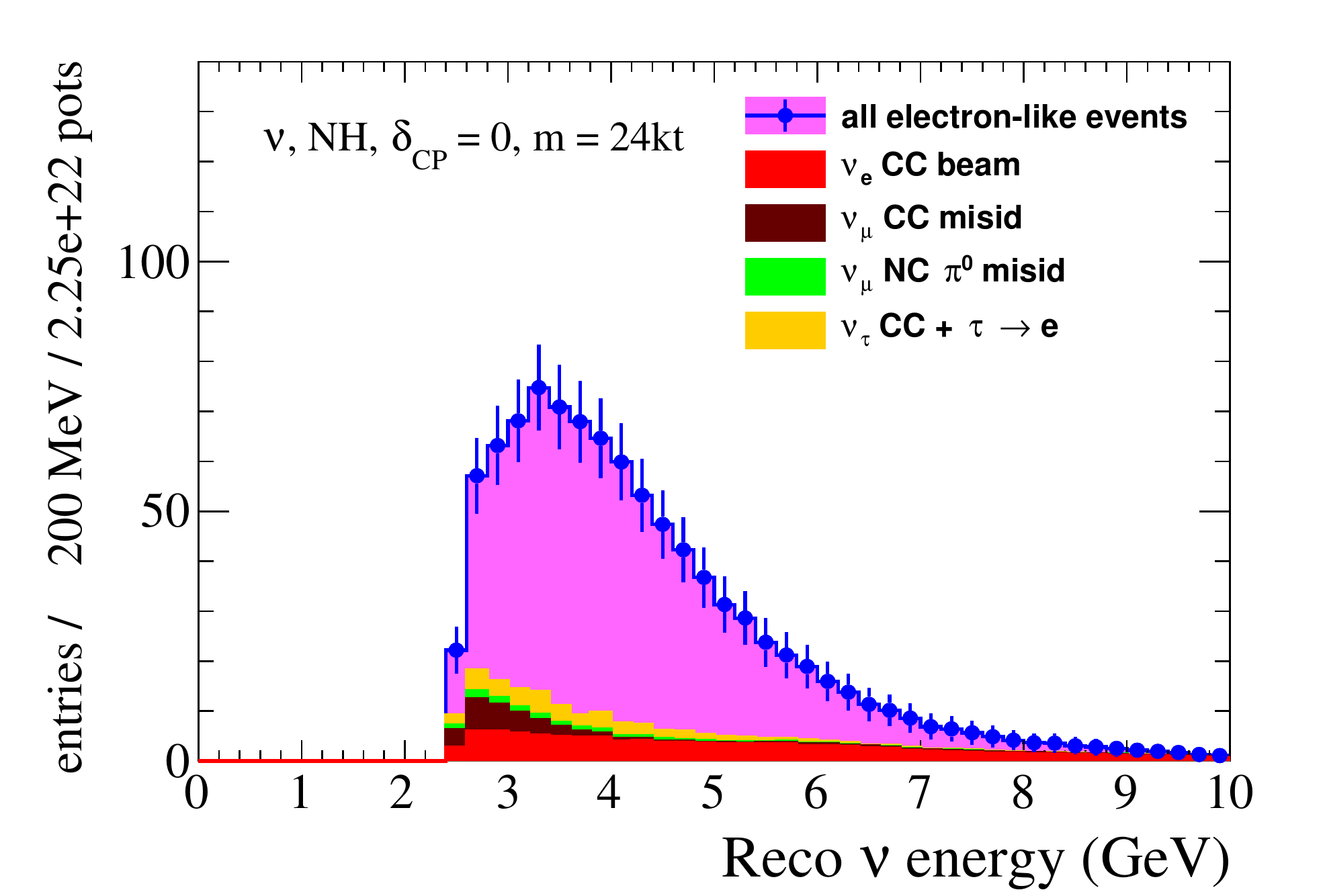}
\includegraphics[width=\the\figwidth]{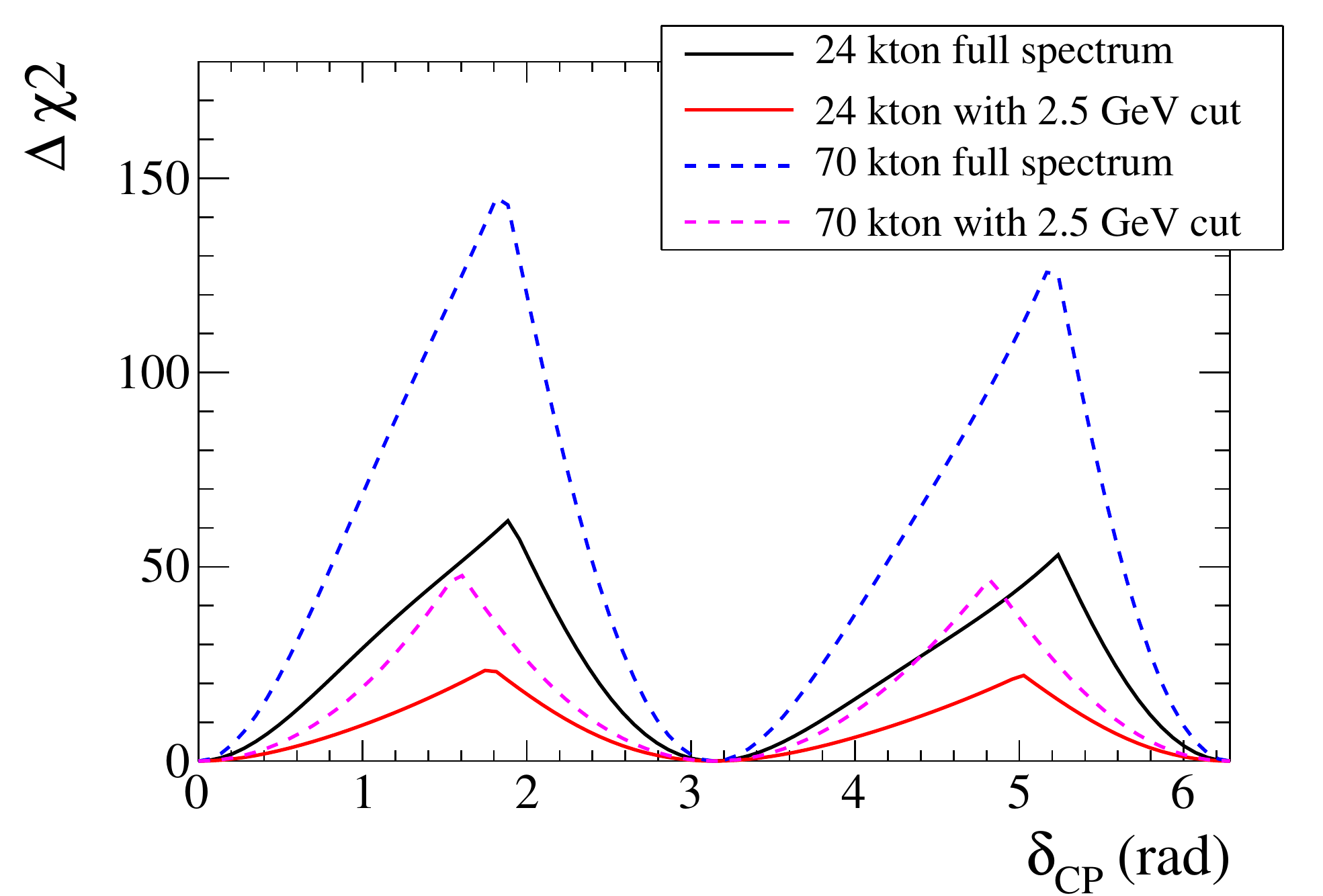}
\caption{Left: Expected distribution of e-like events for HPPS-based neutrino beam after a cut on reconstructed energy at 2.5 GeV. Right: Comparison of the sensitivity to CPV with and without the energy cut.}
\label{fig:hpps_2ndmax_ecut}
\end{center}
\end{figure}

As noted already, the long baseline of 2300~km coupled to the optimised neutrino flux allows LBNO to efficiently detect events in the region of the 1st and 2nd oscillation maxima (see Figs.~\ref{fig:nue_eventdist_sps} and~\ref{fig:nue_eventdist_hpps}). Furthermore the contribution of these $\nu_e$ signal events is critical for enhancing the sensitivity of LBNO to CPV. To illustrate this, we have performed the analysis where all the events with reconstructed neutrino energy below 2.5 GeV had been removed from the sample. The resultant event distributions are shown on the left in Figure~\ref{fig:sps_2ndmax_ecut} and Figure~\ref{fig:hpps_2ndmax_ecut} for SPS and HPPS beams. As evident from these figures, the applied energy cut completely removes any information about the 2nd oscillation maximum. Therefore any deterioration observed in the experimental sensitivity to CPV could only be attributed to loss of the knowledge from this region of $L/E$. 

In the case of the SPS beam, the applied cut results in about 5\% loss in the total number of signal $\nu_e$ events. Although this is a relatively small number, the impact these events have on the CPV sensitivity is not negligible as shown in the plot on the right in Figure~\ref{fig:sps_2ndmax_ecut}. In this case, the coverage at $3\sigma$ level decreases from 45\% (63\%) to 34\% (53\%) for the 24 (70) kton detector, while the coverage at $5\sigma$ level for 70 kton detector option is reduced by more than half (from 35\% to 16\%). The effect is even more pronounced in the case of HPPS beam as can be seen in the right panel Figure~\ref{fig:hpps_2ndmax_ecut}. Here, the cut at 2.5 GeV removes approximately 17\% of signal events. At the same time it leads to a significant loss in the discovery potential with the fraction of $\delta_{CP}$ parameter space covered at $5\sigma$ reduced from 43\% to 0 (65\% to 28\%) for 24 (70) kton detector. These results clearly illustrates that LBNO is not only capable of observing events around 2nd oscillation maximum with appropriate beam optimisation, but also that these events play an important role in determination of the $\delta_{CP}$ phase.

%%%%%%%%%%%%%%%%%%%%%%%%%%%%%%%%%%%%%%%%%%%%%%%%%%%%%%%%%%%%%%%%%%%%%
%             Conclusion
%%%%%%%%%%%%%%%%%%%%%%%%%%%%%%%%%%%%%%%%%%%%%%%%%%%%%%%%%%%%%%%%%%%%%
\section{Conclusions}
\label{sec:concl}

In this paper we have demonstrated that the LBNO setup with a baseline of 2300~km and with
neutrino fluxes optimised using a genetic algorithm, yields an optimal sensitivity 
to the CP-violating $\delta_{CP}$.
With a $\sim 20$~kton double-phase liquid argon detector in Pyh\"{a}salmi and a conventional neutrino beam based on the CERN SPS, the experiment is able to provide a fully conclusive resolution of the neutrino MH problem and has sensitivity to measure CPV to at least the $3\sigma$ level for 45\% of the possible values of $\delta_{CP}$. Further addition of a 50~kton  detector allows to discovery CPV at the level of $3\sigma$ for 63\% of the parameter phase-space. In addition, one would have the sensitivity to reach at least $5\sigma$ level for 35\% of $\delta_{CP}$ values.  Ultimately, with the high intensity neutrino beam delivered by 2MW HPPS, one reaches an impressive sensitivity at the $5\sigma$ ($3\sigma$) level for 65\% (80\%) of the $\delta_{CP}$ phase-space.

The unique and key features of the LBNO setup at 2300~km are its ability to measure the CPV effects not only at the 1st but also the 2nd oscillation maximum. Such spectral information collected over the broad energy window will precisely map the $L/E$ behaviour of the neutrino and antineutrino appearance probabilities which will be a mandatory requirement to fully test the 3-neutrino paradigm.

\section*{Acknowledgments}
We are grateful to the European Commission for the financial support of the project through the FP7 Design Studies LAGUNA (Project Number 212343) and LAGUNA-LBNO (Project Number 284518). We would also like to acknowledge the financial supports of the Lyon Institute of Origins LabEx program (ANR-10-LABEX-66). In addition, participation of individual researchers and institutions has been further supported by funds from ERC (FP7).


\begin{thebibliography}{99}

%\cite{Stahl:2012exa}
\bibitem{Stahl:2012exa} 
 A.~Stahl {\it et al.}, 
Expression of Interest for a 
``Very long baseline neutrino oscillation experiment (LBNO)''; 
CERN SPSC, June, 2012. (CERN-SPSC-2012-021 (SPSC-EOI-007)).
  %%CITATION = CERN-SPSC-2012-021;%%

%\cite{::2013kaa}
\bibitem{Agarwalla::2013kaa} 
  S.~K.~Agarwalla {\it et al.}  [LAGUNA-LBNO Collaboration],
  ``The mass-hierarchy and CP-violation discovery reach of the LBNO long-baseline neutrino experiment,''
  JHEP {\bf 1405}, 094 (2014)
  [arXiv:1312.6520 [hep-ph]].
  %%CITATION = ARXIV:1312.6520;%%
  %19 citations counted in INSPIRE as of 20 Oct 2014

%\cite{Patzak:2012rz}
\bibitem{Patzak:2012rz} 
  T.~Patzak [LAGUNA-LBNO Collaboration],
  ``LAGUNA and LAGUNA-LBNO: Future megaton neutrino detectors in Europe,''
  Nucl.\ Instrum.\ Meth.\ A {\bf 695}, 184 (2012).
  %%CITATION = NUIMA,A695,184;%%
  
  %\cite{Rubbia:2013zqa}
\bibitem{Rubbia:2013zqa} 
  A.~Rubbia,
  ``LAGUNA-LBNO: Design of an underground neutrino observatory coupled to long baseline neutrino beams from CERN,''
  J.\ Phys.\ Conf.\ Ser.\  {\bf 408}, 012006 (2013).
  %%CITATION = 00462,408,012006;%%
  %1 citations counted in INSPIRE as of 20 Oct 2014
    
    %\cite{Rubbia:2009md}
\bibitem{Rubbia:2009md}
  A.~Rubbia,
  ``Underground Neutrino Detectors for Particle and Astroparticle Science: The Giant Liquid Argon Charge Imaging ExpeRiment (GLACIER),''
  J.\ Phys.\ Conf.\ Ser.\  {\bf 171} (2009) 012020
  [arXiv:0908.1286 [hep-ph]].
  %%CITATION = ARXIV:0908.1286;%%
  %60 citations counted in INSPIRE as of 22 Oct 2014
  
   %\cite{Adams:2013qkq}
\bibitem{Adams:2013qkq}
  C.~Adams {\it et al.}  [LBNE Collaboration],
  ``The Long-Baseline Neutrino Experiment: Exploring Fundamental Symmetries of the Universe,''
  arXiv:1307.7335 [hep-ex].
  %%CITATION = ARXIV:1307.7335;%%
  %88 citations counted in INSPIRE as of 22 Oct 2014
  
  %\cite{Bass:2013vcg}
\bibitem{Bass:2013vcg}
  M.~Bass {\it et al.}  [LBNE Collaboration],
  ``Baseline optimisation for the measurement of CP violation and mass hierarchy in a long-baseline neutrino oscillation experiment,''
  arXiv:1311.0212 [hep-ex].
  %%CITATION = ARXIV:1311.0212;%%
  %6 citations counted in INSPIRE as of 22 Oct 2014
  
 %\cite{Kearns:2013lea}
\bibitem{Kearns:2013lea}
  E.~Kearns {\it et al.}  [Hyper-Kamiokande Working Group Collaboration],
  ``Hyper-Kamiokande Physics Opportunities,''
  arXiv:1309.0184 [hep-ex].
  %%CITATION = ARXIV:1309.0184;%%
  %12 citations counted in INSPIRE as of 22 Oct 2014
   
  %\cite{Barger:2001yr}
\bibitem{Barger:2001yr}
  V.~Barger, D.~Marfatia and K.~Whisnant,
  ``Breaking eight fold degeneracies in neutrino CP violation, mixing, and mass hierarchy,''
  Phys.\ Rev.\ D {\bf 65} (2002) 073023
  [hep-ph/0112119].
  %%CITATION = HEP-PH/0112119;%%
  %359 citations counted in INSPIRE as of 22 Oct 2014
  
   %\cite{Coloma:2011pg}
\bibitem{Coloma:2011pg}
  P.~Coloma and E.~Fernandez-Martinez,
  ``optimisation of neutrino oscillation facilities for large $\theta_{13}$,''
  JHEP {\bf 1204} (2012) 089
  [arXiv:1110.4583 [hep-ph]].
  %%CITATION = ARXIV:1110.4583;%%
  %10 citations counted in INSPIRE as of 22 Oct 2014
 
 \bibitem{ipac13_hpps} Y.~Papaphilippou et al., "Design options of a High-Power Proton Synchrotron for LAGUNA-LBNO'', THPWO081, IPAC13 proceedings.

  %\cite{Ferrari:2005zk}
\bibitem{Ferrari:2005zk} 
  A.~Ferrari, P.~R.~Sala, A.~Fasso and J.~Ranft,
  ``FLUKA: A multi-particle transport code (Program version 2005),''
  CERN-2005-010, SLAC-R-773, INFN-TC-05-11.
  %%CITATION = CERN-2005-010, SLAC-R-773, INFN-TC-05-11;%%
  %263 citations counted in INSPIRE as of 20 Oct 2014

\bibitem{DEAP_JMLR2012}
F\'elix-Antoine Fortin, Fran\c{c}ois-Michel {De Rainville}, Marc-Andr\'e Gardner, Marc Parizeau, Christian Gagn\'e,
``{DEAP}: Evolutionary Algorithms Made Easy,''
Journal of Machine Learning Research {\bf 13} (2012) 2171-2175.

%\cite{Huber:2004ka}
\bibitem{Huber:2004ka} 
  P.~Huber, M.~Lindner and W.~Winter,
  ``Simulation of long-baseline neutrino oscillation experiments with GLoBES (General Long Baseline Experiment Simulator),''
  Comput.\ Phys.\ Commun.\  {\bf 167}, 195 (2005)
  [hep-ph/0407333].
  %%CITATION = HEP-PH/0407333;%%
  %273 citations counted in INSPIRE as of 20 Oct 2014
  
  %\cite{Abe:2014ugx}
\bibitem{Abe:2014ugx} 
  K.~Abe {\it et al.}  [T2K Collaboration],
  ``Precise Measurement of the Neutrino Mixing Parameter $\theta_{23}$ from Muon Neutrino Disappearance in an Off-axis Beam,''
  Phys.\ Rev.\ Lett.\  {\bf 112}, 181801 (2014)
  [arXiv:1403.1532 [hep-ex]].
  %%CITATION = ARXIV:1403.1532;%%
  %29 citations counted in INSPIRE as of 20 Oct 2014
  
  %\cite{GonzalezGarcia:2012sz}
\bibitem{GonzalezGarcia:2012sz} 
  M.~C.~Gonzalez-Garcia, M.~Maltoni, J.~Salvado and T.~Schwetz,
  ``Global fit to three neutrino mixing: critical look at present precision,''
  JHEP {\bf 1212}, 123 (2012)
  [arXiv:1209.3023 [hep-ph]].
  %%CITATION = ARXIV:1209.3023;%%
  %414 citations counted in INSPIRE as of 20 Oct 2014
  
\end{thebibliography}
\end{document}